\documentclass{article}
\usepackage{graphicx}  
\usepackage{amsmath}   
\usepackage[compress]{cite}
\usepackage{amssymb}   
\usepackage{bm} 
\usepackage{dcolumn}
\usepackage{color}
\usepackage{mathrsfs}
\usepackage[utf8]{inputenc}
\usepackage{amsfonts}
\usepackage{varioref}
\usepackage{color}
\usepackage{subfig}
\usepackage{float}
\usepackage{booktabs,multirow,array}
\usepackage[table]{xcolor}
\usepackage{float}
\RequirePackage[colorlinks,citecolor=blue,urlcolor=magenta,linkcolor=blue]{hyperref}
\def\EGB{Einstein-Gauss-Bonnet}

\def\scc{strong cosmic censorship}

\def\gr{general relativity}

\def\RNDS{Reissner-Nordstr\"{o}m-de Sitter }
\def\cch{cosmic censorship hypothesis}

\def\ch{Cauchy horizon}

\def\qnm{quasi-normal }
\def\PS{photon sphere }
\def\NE{near extremal }
\def\DS{de Sitter }

\labelformat{section}{Section #1} 
\labelformat{subsection}{Section #1} 
\labelformat{subsubsection}{Section #1}
\labelformat{subsubsubsection}{Section #1}
\labelformat{equation}{Eq.~(#1)} \labelformat{figure}{Fig.~#1} 
\labelformat{subfigure}{Fig.~\thefigure#1} 
\labelformat{table}{Table.~#1} 
\labelformat{appendix}{Appendix #1}
\title{\textbf{Quasinormal modes and Strong Cosmic Censorship in the regularised 4D Einstein-Gauss-Bonnet gravity}}
\author{Akash K Mishra\footnote{akash.mishra@iitgn.ac.in}$~^{1}$ \\
{\small{$^{1}$ Indian Institute of Technology, Gandhinagar-382355, Gujarat, India}}\\
}
\date{ }  
\begin{document}
 
\maketitle
\begin{abstract}
The fate of strong cosmic censorship is ultimately linked to the extendibility of perturbation across the Cauchy Horizon and known to be violated in the near extremal region of a charged de Sitter black hole. Similar violations can also be realized in higher curvature theories, with the strength of violation becoming stronger as compared to general relativity. In this work, we extend this analysis further to study the validity of strong cosmic censorship conjecture in the context of the regularised four-dimensional Einstein Gauss-Bonnet theory with respect to both scalar and electromagnetic perturbation. We also study the late time tails of scalar fields.
\end{abstract}
\section{Introduction and Motivation}
Predictability is one of the essential features of any well behaved physical theory of nature. Any physical theory is said to be predictable or deterministic if the future of any associated physical entity can be uniquely determined from regular initial data given on a Cauchy hypersurface. For general relativity, this is ensured by the strong cosmic censorship conjecture implying general relativity to be deterministic in nature. The conjecture in the most general form is essentially the statement that all physically reasonable solutions of Einstein's equation are globally hyperbolic\cite{Wald:1984rg,hawking_ellis_1973}. 
Mathematically, this can be captured by the statement that the domain of dependence of the Cauchy hypersurface is the entire spacetime manifold. However, the existence of Cauchy horizon in various black hole solutions of Einstein's equation may lead to a possible violation of the \scc\, since a \ch\ represents the boundary of maximal evolution of initial data. Therefore, the classical fate of any observer beyond the \ch\ can not be uniquely determined in general relativity by evolving initial data on a Cauchy hypersurface. A situation such as this, questions the credibility of general relativity as a deterministic theory and must be addressed.
One possible resolution to this paradox in the context of asymptotically flat spacetime is the unstable nature of \ch, a surface of infinite blueshift. This phenomenon is known as the mass inflation \cite{PhysRevD.41.1796,PhysRevLett.67.789,poisson_2004,Bhattacharjee:2016zof}. The exponential growth, i.e., $\Phi\sim e^{\kappa_{-}u}$, with $\kappa_{-}$ being the surface gravity of the \ch, dominates the late-time power-law decay of any perturbation and further resulting into the formation of a  curvature singularity at the \ch. This suggests that it is impossible to extend any perturbation across the \ch\ with $C^2$ spacetime metric. However, a divergence in curvature at the \ch\ does not necessarily imply that the spacetime can not be extended beyond. Several solutions of Einstein's equation, namely the \textit{weak solution} exists with spacetime metric having regularity lower than $C^2$. This further leads to a revised version of the \scc\ as formulated by Christodoulou, which state that, it is impossible to extend the spacetime beyond \ch\ with locally square integrable Christoffel connection, i.e., $\Gamma^{a}_{bc}\in L^2_{loc}$\cite{Christodoulou:2008nj}.
Although the Christodoulou version of \scc\ trivially holds for asymptotically flat black holes, the situation is drastically different in the case of asymptotically de Sitter spacetimes. The presence of cosmological horizon leads to an exponentially decaying behaviour of perturbations at a late time, i.e., $\Phi\sim e^{-\alpha u}$, where $\alpha = -Im(\omega)$ is the spectral gap related to the longest-lived quasi-normal mode. Therefore, it is indeed the relative strength between the late time redshift effect and near \ch\ blueshift effect($\beta=\alpha/\kappa_{-}$) that decides the fate of \scc\ in the presence of a positive cosmological constant. 
\\

Recent interest in the subject stems from the work of Cardoso et al.  \cite{PhysRevLett.120.031103}, where a violation to the Christodoulou version of \scc\ in the near extremal region has been reported in the context of a massless scalar perturbation on a \RNDS\ background. 
Such an approach has been followed recently by several authors to study the validity of the \scc\ in various black hole spacetimes\cite{Mo:2018nnu,Dias:2018ufh,Ge:2018vjq,Rahman:2018oso,Destounis:2018qnb,Gwak:2018rba,Gim:2019rkl,Liu:2019lon,Rahman:2019uwf,Guo:2019tjy,Dias:2019ery,Liu:2019rbq,Cardoso:2018nvb,Dias:2018etb,Gan:2019jac,Destounis:2019omd,Rahman:2020guv}. Surprisingly, no such violation occurs in Kerr-de Sitter black holes\cite{Dias:2018ynt}.
Very recently, this analysis had been generalized to study the effects of higher curvature terms on the validity of \scc\cite{Mishra:2019ged,Gan:2019ibg}. Particularly in Ref.\cite{Mishra:2019ged}, we have explicitly demonstrated that a violation of the conjecture occurs in the near extremal region by considering the propagation of a massless scalar perturbation in five-dimensional \EGB\  and pure lovelock theories. Another central result of Ref.\cite{Mishra:2019ged}is the realization that the violation of \scc\  becomes stronger with the increasing strength of the higher curvature coupling constant. The \EGB\ theory and its full Lovelock generalization represent the low energy effective stringy correction over the Einstein-Hilbert action with the field equation containing the second derivative of the metric\cite{Lovelock:1971yv,ZWIEBACH1985315,PhysRevLett.55.2656,ZUMINO1986109,PADMANABHAN2013115,Chakraborty:2015wma}. The Gauss-Bonnet term turns out to be a total derivative in $D=4$. However, in $D>4$ the Gauss-Bonnet term contributes to the gravitational dynamics giving rise to various non-trivial results that have been extensively studied in the past few decades. Recently, the \EGB\ theory of gravity in $D=4$ has been reformulated as the $D\to4$ limit of a higher dimensional theory after rescaling the coupling constant as $\alpha\to \alpha/(D-4)$\cite{Glavan:2019inb}. The newly formulated four-dimensional Einstein-Gauss-Bonnet theory bypasses the Lovelock theorem and also is free of Ostrogradsky instabilities.\\

Several objections have been raised regarding the validity of the four dimensional theory\cite{Gurses:2020ofy,Shu:2020cjw,Ai:2020peo} and the regularization scheme used in Ref\cite{Glavan:2019inb}, with all leading to the conclusion that, no pure four dimensional Einstein Gauss Bonnet theory exists. To circumvent this problem, several regularization schemes have been proposed with well defined $D\to 4$ limit\cite{Lu:2020iav,Kobayashi:2020wqy,Fernandes:2020nbq,Hennigar:2020lsl}. Such regularized four dimensional theories are typically scalar tensor theories of Horndiski type with an additional scalar degree of freedom.\\

These theories admits same spherically symmetric black hole solution in four dimensions as proposed in\cite{Glavan:2019inb} and provides a natural alternative to black hole solutions of general relativity\cite{Lu:2020iav,Hennigar:2020lsl,1801153}. This has stimulated a series of recent research work to study various properties of the novel four dimensional black hole solution in the \EGB\ theory of gravity\cite{Konoplya:2020bxa,Fernandes:2020rpa,Guo:2020zmf,Konoplya:2020qqh,Kumar:2020owy,Ghosh:2020vpc,Lu:2020iav,Ghosh:2020syx,Konoplya:2020juj,Churilova:2020aca,Zhang:2020qam,Wei:2020poh,Hegde:2020xlv,Zhang:2020qew,Wei:2020ght,HosseiniMansoori:2020yfj}. In this work, we extend our previous analysis of exploring the violation of strong cosmic censorship in the presence of higher curvature term to the regularized $4D$ \EGB\ gravity. In particular, we are intersted in understanding the effect of Gauss-Bonnet coupling constant on the violation of \scc. Another interesting aspect of this work is to understand whether the violation of \scc\ in four dimension is stronger or weaker as compared to the five dimensional theory.
Our study includes the case of both scalar and electromagnetic perturbations.
\\

As previously discussed, the fate of \scc\ conjecture is ultimately governed the relative ratio between the blueshift effect at the \ch\ and the late time decaying behaviour of the perturbation, i.e., $\beta$. 
In order to obtain the bound on $\beta$, let us consider a test scalar field perturbation on a spherically symmetric static black hole background. The dynamics of the scalar field is governed by the Klein Gordon equation $\square\Phi = 0$. The spacetime symmetry allows one to decompose the scalar field as $\Phi(t,r,\theta,\phi)=e^{-i\omega t}R(r)Y_{l m}(\theta,\phi)$, where $Y_{l m}$ represents the spherical harmonics and $\omega$ is the quasi-normal frequency. The radial function $R(r)$ satisfies a second-order differential equation with two independent solutions given by,
\begin{align}
\Phi^{(1)}(t,r,\theta,\phi)=e^{-i\omega u}R^{(1)}(r)Y_{l m}(\theta,\phi);\qquad \Phi^{(2)}(t,r,\theta,\phi)=e^{-i\omega u}R^{(2)}(r)(r-r_{-})^{i\omega_{n}/\kappa_{-}}Y_{l m}(\theta,\phi)
\end{align}
The integral of the kinetic term of the scalar field turns out to be proportional to $(r-r_{-})^{2\beta -1}$. As a result, the scalar field is regular at the \ch\ and can be extended beyond if $\beta>1/2$.
The case of electromagnetic perturbation turns out to be identical to that of the scalar perturbation. However, the extendibility of gravitational perturbation across the \ch\ turns out to be more subtle in higher curvature theories, and one requires a stronger regularity condition at the \ch. For a detailed discussion, refer to\cite{Mishra:2019ged,Destounis:2019omd}. Also, note that the late time exponential decay of perturbation in de Sitter space is one of the necessary ingredients for the definition of $\beta$ to make sense. In the case of Einstein's gravity and higher dimensional \EGB\ theory, such late time tail has been extensively studied for various black hole spacetime in de Sitter space. 
Since the late time behaviour depends on the asymptotic structure of the effective potential, it is reasonable to expect identical behaviour for the four dimensional \EGB\ theory as well. However, such expectation should be supported by strong numerical analysis. We establish this result by performing the time domain evolution for both scalar and electromagnetic perturbations.\\

The remainder of the article is organised as follows: In \ref{Section 2}, we start by briefly reviewing spherically symmetric charged black hole solutions in the regularized $4D$ \EGB\ theory of gravity. Subsequently, in \ref{Section 3}, we discuss the various dominating family of modes relevant for the analysis of \scc. \ref{Section 4} is devoted to numerical analysis: firstly we study the late time dynamics of scalar perturbation and then compute quasinormal modes, secondly we demonstrate the violation of \scc\ in the context of scalar and electromagnetic perturbations. In \ref{Section 5}, we summarize our results and discuss possible future outlooks.

\section{Black holes in the regularised $4D$ Einstein Gauss-Bonnet gravity}\label{Section 2}
Higher curvature terms naturally occur in various occasion as the low energy effective action of superstring theories. General Relativity is a perturbatively non-renormalizable theory, and in order to make sense, it must be supplemented by these higher curvature corrections over the Einstein-Hilbert action in the regime of strong gravity. Among several class of such higher curvature terms, the Lovelock corrections play a very special role. The field equation contains at most up to the second derivative of the metric, and the gravitational dynamics is free from Ostrogradsky instabilities. Inclusion of such terms to the gravitational action leads to various interesting results that are distinct from general relativity. This motivates one to explore various aspects of black holes in such theories.
The third-order term in the Lovelock polynomial represents the Gauss-Bonnet correction, which turns out to be a total divergence in $D=4$ and does not contribute to the gravitational dynamics. However, recently in Ref.\cite{Glavan:2019inb}, Glavan and Lin have formulated a \EGB\ theory in four dimensions as the $D\to4$ limit of a higher dimensional theory after the coupling constant rescaled as $\alpha \to \alpha/(D-4)$. The action for the novel \EGB\ theory with the rescaled coupling is of the form,
\begin{equation}
S = \frac{1}{16\pi}\int d^{D}x\sqrt{-g}\left[-2\Lambda + R + \frac{\alpha}{D-4}\left(R_{abcd}R^{abcd}- 4 R_{ab}^{ab} + R^2 \right) - 4\pi F_{ab}F^{ab} \right]
\end{equation}
Here $\Lambda$ represents the cosmological constant, $R_{abcd}$ is the Riemann tensor in $D-$dimensions and $g$ is the determinant of the metric. The Gauss-Bonnet coupling constant is denoted by $\alpha$, and $F_{ab}$ is the electromagnetic field tensor. The four-dimensional novel \EGB\ theory is then obtained by varying the action and imposing the $D\to4$ limit at the level of field equation. However, the theory has faced several criticism regarding its validity as a consistent four dimensional theory. As illustrated in Ref\cite{Gurses:2020ofy}, the main problem has to do with taking $D\to 4$ limit of a higher dimensional theory. In this procedure, either there always exists a higher dimensional part of the field equation or one looses the Bianchi identity. To resolve this issue, several regularized theory have been proposed with consistent four dimensional limit. Particularly, Ref\cite{Lu:2020iav}, used a Kaluza-Klein reduction to compactify the $D-$dimensional theory on a $(D-4)-$ dimensional maximally symmetric spacetime and then rescaling the coupling constant to obtain a well defined four dimensional theory. The resulting theory is a scalar tensor theory of Horndiski type.  Another way to obtain a consistent $D\to 4$ limit without reference to any higher dimensional spacetime has been discussed in \cite{Hennigar:2020lsl}. The central result of these papers is the realization that, it is indeed possible to have a consistent four dimensional description of \EGB\ theory of gravity, but with an additional scalar degree of freedom.  
Interestingly, both of these regularized theories admits spherically symmetric charged black hole solutions of the form\cite{Fernandes:2020rpa,Ghosh:2020syx},

\begin{equation}\label{metric}
ds^2=-f(r)dt^2 + f(r)^{-1} dr^2 + r^2 d\Omega^2
\end{equation}
where the metric coefficient $f(r)$ is given by,
\begin{equation}
f(r) = 1+\frac{r^2}{2\alpha}\left[1- \sqrt{1+4\alpha\left(\frac{2M}{r^3}-\frac{Q^2}{r^4}+\frac{\Lambda}{3}\right)}\right]
\end{equation}
The parameter $M$ and $Q$ represents the ADM mass and $U(1)$ charge of the black hole respectively. In the limit $\alpha\to 0$, one obtains the four-dimensional \RNDS\ black hole. Further, we consider only the case of positive Gauss-Bonnet coupling($\alpha>0$) constant. Interestingly, similar black hole solution was also found earlier in the context of semi-classical gravity with conformal anamoly\cite{Cai:2009ua,Cai:2014jea} and third order regularized Lovelock gravity\cite{Casalino:2020kbt}.
The $4D$ \EGB\ charged de Sitter black hole admits three horizon namely the event horizon($r_{+}$), the \ch($r_{-}$) and the cosmological horizon($r_{c}$) given by the solution of the equation $f(r)=0$. The existence of three positive roots of the horizon equation, which is essential for our analysis is ensured by the Descarte rule of sign.

\section{Quasinormal frequency and dominating family of modes: scalar and electromagnetic perturbations}\label{Section 3}
One of the central themes of research in contemporary gravitational physics is the stability analysis of black holes under small perturbation by computing the quasi-normal mode spectrum. Quasinormal modes are eigenfrequency of the corresponding perturbation equation with respect to a special set of boundary condition, namely the black hole boundary conditions. Let us start by studying a scalar and electromagnetic perturbation over the novel $4D$ \EGB\ black hole background, which are governed by the following equations respectively\cite{Berti:2009kk,Konoplya:2011qq},
\begin{align}
\frac{1}{\sqrt{-g}}\partial_\mu\left(\sqrt{-g}g^{\mu\nu}\partial_\nu \Phi\right)=0 ~~~ \text{and}~~~\frac{1}{\sqrt{-g}}\partial_\mu\left(\sqrt{-g}g^{\rho\nu}g^{\sigma\mu}\partial_\nu F_{\rho\sigma}\right)=0
\end{align}
which after the separation of variables in terms of spherical harmonics takes the form of an eigenvalue equation with an effective potential.
\begin{align}\label{mastereqn}
\left(\frac{\partial^2}{\partial r_{*}^2}+\omega^2-V_{\rm s}(r)\right)\Psi_s(r)=0; ~~ V_{\rm s}(r) = f(r)\left(\frac{\ell(\ell+1)}{r^2}+\frac{(1-s^2)f'(r)}{r}\right)
\end{align}
Because of the complicated structure of the effective potential, it is challenging to solve for the eigenfrequency modes analytically, and hence such systems are often dealt with various numerical techniques. However, for certain limiting cases, the effective potential attains a simplified form, and one can solve the eigenvalue equation analytically. One such case is the limit of large angular momentum($\ell\to\infty$), namely the eikonal approximation. In this approximation, the real and imaginary part of the quasi-normal frequency are directly related to the angular velocity($\Omega_{\rm ph}$) and the instability time scale\, namely the Lyapunov exponent($\lambda_{\rm ph}$) associated with the photon sphere around the black hole\cite{PhysRevD.79.064016,PhysRevD.80.064004,PhysRevD.31.290,Cornish:2003ig,Bombelli_1992,PhysRevLett.52.1361,Konoplya:2017wot,Konoplya:2019hml},
\begin{equation}\label{imomega}
\omega_{n}=\Omega_{\rm ph}\ell-i\left(n+\frac{1}{2}\right)\lambda_{\rm ph}~,
\end{equation}
where the Lyapunov exponent is given by,
\begin{equation}\label{lyapunov}
\lambda_{\rm ph} = \sqrt{\frac{f(r_{\rm ph})}{2}\left(\frac{2f(r_{\rm ph})}{r_{\rm ph}^{2}}-f''(r_{\rm ph})\right)}
\end{equation}
Here $`n$' represents the overtone number and $r_{\rm ph}$ is the radius of the photon sphere. For obvious reason, the quasi-normal frequencies in the eikonal approximation are referred to be the photon sphere modes.
Note that, in the $\ell\to\infty$ limit, the effective potential is dominated by the $\ell(\ell+1)$ term and the effect of spin turns out to be negligible. Therefore, one should expect to have similar quasi-normal frequencies for both scalar and electromagnetic perturbation in the eikonal limit. This can be clearly realized from the numerical data presented in \ref{table-QNM-GB_4d_scalar} and \ref{table-QNM-GB_4d_em} of the subsequent sections.\\
 
Our second family of relevant modes are the de Sitter modes which dominate the quasi-normal spectrum in the limit $\Lambda\to0$ when the \cch lies far away from the event horizon. The associated quasi-normal frequency for scalar perturbation in this limit turns out to be purely imaginary since they can be considered as a deformation of pure de Sitter modes and given by\cite{PhysRevD.70.064024,LopezOrtega:2012vi,PhysRevD.66.104018,Abdalla:2005hu},
\begin{equation}\label{qnmds}
\omega _{n,\textrm{dS}}=-i\left(\ell+2n\right) \kappa _{\rm c}
\end{equation} 
The dominating mode corresponds to $\ell=1$ and $n=0$, with $\kappa_{c}$ being the surface gravity of the cosmological horizon. Now, by following an identical line of calculation as in Ref.\cite{Natario:2004jd}, we obtain an expression for the pure de Sitter modes with respect to electromagnetic perturbation. To achieve this, we recast the potential for electromagnetic perturbation in \ref{mastereqn} as,
\begin{align}\label{empotential}
 V_{\rm em}(r) = (1-\lambda r^2)\left(\frac{\ell(\ell+1)}{r^2}\right) ~~~ \text{where} ~~~ \lambda= \frac{1}{2\alpha}\left[\sqrt{1+\frac{4\alpha \Lambda}{3}}-1\right]
\end{align}
Upon further substitution, $\Phi(r)=r^{\ell+1}(1-\lambda r^2)^{(i \tilde{\omega}/2)}\psi(r)$, the electromagnetic perturbation equation in \ref{mastereqn} takes the following form,

\begin{align}\label{em_new}
r(1-\lambda r^2)\frac{d^2\psi}{dr^2} +2\left[1+(1-\lambda r^2)\ell -(2+ i\tilde{\omega})\lambda r^2 \right]\frac{d\psi}{dr}-\lambda r\left[2+\ell^2+\ell(3+2i\tilde{\omega})+3 i \tilde{\omega}^2 -\tilde{\omega}^2\right]\psi =0
\end{align}
where $\tilde{\omega}=(\frac{\omega}{\sqrt{\lambda}})$. The above equation attains exact solution which is given in terms of hypergeometric functions as follows,
\begin{align}
\Phi(r) &= C_{1}\,\times  {}_{2}F_{1}\left[\frac{2+2\ell+2 i\tilde{\omega}}{4},\frac{4+2\ell +2 i \tilde{\omega}}{4}, \frac{3+2\ell}{2}\Bigg|\lambda r^2 \right] r^{\ell+1}(1-\lambda r^2)^\frac{i \tilde{\omega}}{2}\nonumber\\&
+ C_{2}\,\times  {}_{2}F_{1}\left[\frac{2 i \tilde{\omega}-2\ell}{4},\frac{2-2\ell +2 i \tilde{\omega}}{4}, \frac{1-2\ell}{2}\Bigg|\lambda r^2 \right] r^{-\ell}(1-\lambda r^2)^\frac{i \tilde{\omega}}{2}
\end{align}
Now, in order to extract the quasi-normal modes spectrum, we impose the black hole boundary conditions. At the origin, we demand the solution to be regular which requires $C_2=0$ and we are left with only one term. Using the identities of hypergeometric functions, it is useful to write the above solution in the following form,

\begin{align}
& {}_{2}F_{1}\left[\frac{2+2\ell+2 i\tilde{\omega}}{4},\frac{4+2\ell +2 i \tilde{\omega}}{4}, \frac{3+2\ell}{2}\Bigg|\lambda r^2 \right] (1-\lambda r^2)^\frac{i \tilde{\omega}}{2}\nonumber\\&
= \frac{(1-\lambda r^2)^{(-i \tilde{\omega}/2)}\,\Gamma[i \tilde{\omega}]\, \Gamma[\frac{3+2\ell}{2}]}{\Gamma[\frac{2+2\ell+2 i \tilde{\omega}}{4}]\,\Gamma[\frac{4+2\ell+2 i \tilde{\omega}}{4}]} \times {}_{2}F_{1}\left[\frac{4+2\ell-2 i\tilde{\omega}}{4},\frac{2+2\ell -2 i \tilde{\omega}}{4},1-i\tilde{\omega}\Bigg|1-\lambda r^2 \right] \nonumber\\&
+ \frac{(1-\lambda r^2)^{(i \tilde{\omega}/2)}\,\Gamma[-i \tilde{\omega}]\, \Gamma[\frac{3+2\ell}{2}]}{\Gamma[\frac{4+2\ell-2 i \tilde{\omega}}{4}]\,\Gamma[\frac{2+2\ell-2 i \tilde{\omega}}{4}]} \times {}_{2}F_{1}\left[\frac{2+2\ell+2 i\tilde{\omega}}{4},\frac{4+2\ell +2 i \tilde{\omega}}{4},1+ i\tilde{\omega}\Bigg|1-\lambda r^2 \right] 
\end{align}
At the cosmological horizon we have $(1-\lambda r^2)\to 0$ and we use the identity ${}_{2}F_{1}[a,b,c|0]=1$. The outgoing mode boundary condition at the cosmological horizon further gives rise to,
\begin{align}
 \frac{\Gamma[i \tilde{\omega}]\, \Gamma[\frac{3+2\ell}{2}]}{\Gamma[\frac{2+2\ell+2 i \tilde{\omega}}{4}]\,\Gamma[\frac{4+2\ell+2 i \tilde{\omega}}{4}]} =0 \Rightarrow \tilde{\omega}_1 = -i(2 n+ \ell + 1) ~~\text{and}~~  \tilde{\omega}_2 = -i(2 n+ \ell + 2) 
\end{align}
The pure de Sitter quasi-normal frequency for electromagnetic perturbation is then given by,
\begin{equation}
\omega_{\rm n, ds}^{\rm em} = -i (2n+\ell +1)\sqrt{\lambda} = -i (2n+\ell +1)\kappa_{c}
\end{equation}
where $\kappa_{c}$ is the surface gravity associated with the cosmological horizon.The dominant mode which is relevant for our analysis corresponds to the choice of $(n=0,\ell=1)$, i.e., $\omega_{\rm ds}^{em}=-2 i \kappa_{c}$. Although the analytical expressions for the photon sphere mode and the de Sitter mode in this section have been obtained using various approximation, they provide excellent agreement with the numerical results obtained in the subsequent sections. A final set of modes relevant to our discussion is the near extremal modes, which dominate the quasi-normal mode spectrum in the limit when the event horizon and the \ch\ approach each other. Unlike the case of general relativity, it is a daunting task to come up with an analytical expression for the near extremal modes in \EGB\ theory. In this article, we shall not attempt to obtain such an expression; rather, we would compute it numerically.

\section{Numerical analysis and validity of strong cosmic censorship}\label{Section 4}
In the previous section, we have discussed the various dominant set of modes that are relevant in the context of \scc. In particular, we obtained an expression for the quasi-normal frequency associated with the de Sitter mode for electromagnetic perturbations.
This section is devoted to the computation of quasinormal modes and studying the validity of \scc\ for scalar(s=0) and electromagnetic perturbations(s=1). Our approach involves the computation of quasi-normal frequencies numerically and then studying the variation of  $\beta$ to see if the conjecture is violated for the theory we consider. It is well known that the quasi-normal mode spectrum of \EGB\ black holes in five dimensions and the violation of \scc\ conjecture strongly depend on the Gauss-Bonnet coupling $\alpha$\cite{Abdalla:2005hu,Konoplya:2004xx,Konoplya:2008ix,Cuyubamba:2016cug,Konoplya:2017ymp,Konoplya:2017lhs,Nojiri:2001aj,Nojiri:2001ae,Mishra:2019ged}. Hence in this context, it would be an interesting exercise to carry out such analysis in the context of the regularized $4D$ \EGB\ theory, which we explore subsequently. We also establish that the late time behaviour obeys an exponential decay, which is one of the essential ingredient in the analysis of \scc\ in de Sitter space.
We compute the quasinormal modes by three different method: firstly by using the time domain evolution, secondly by WKB approximation and finally by the Mathematica package developed in Ref.\cite{Jansen:2017oag}. All the three method shows excellent agreement.

\subsection{Computation of quasinormal frequencies: Time Domain evolution and late time tail}
\begin{figure}[!htp]
\centering
\subfloat{{\includegraphics[scale=0.34]{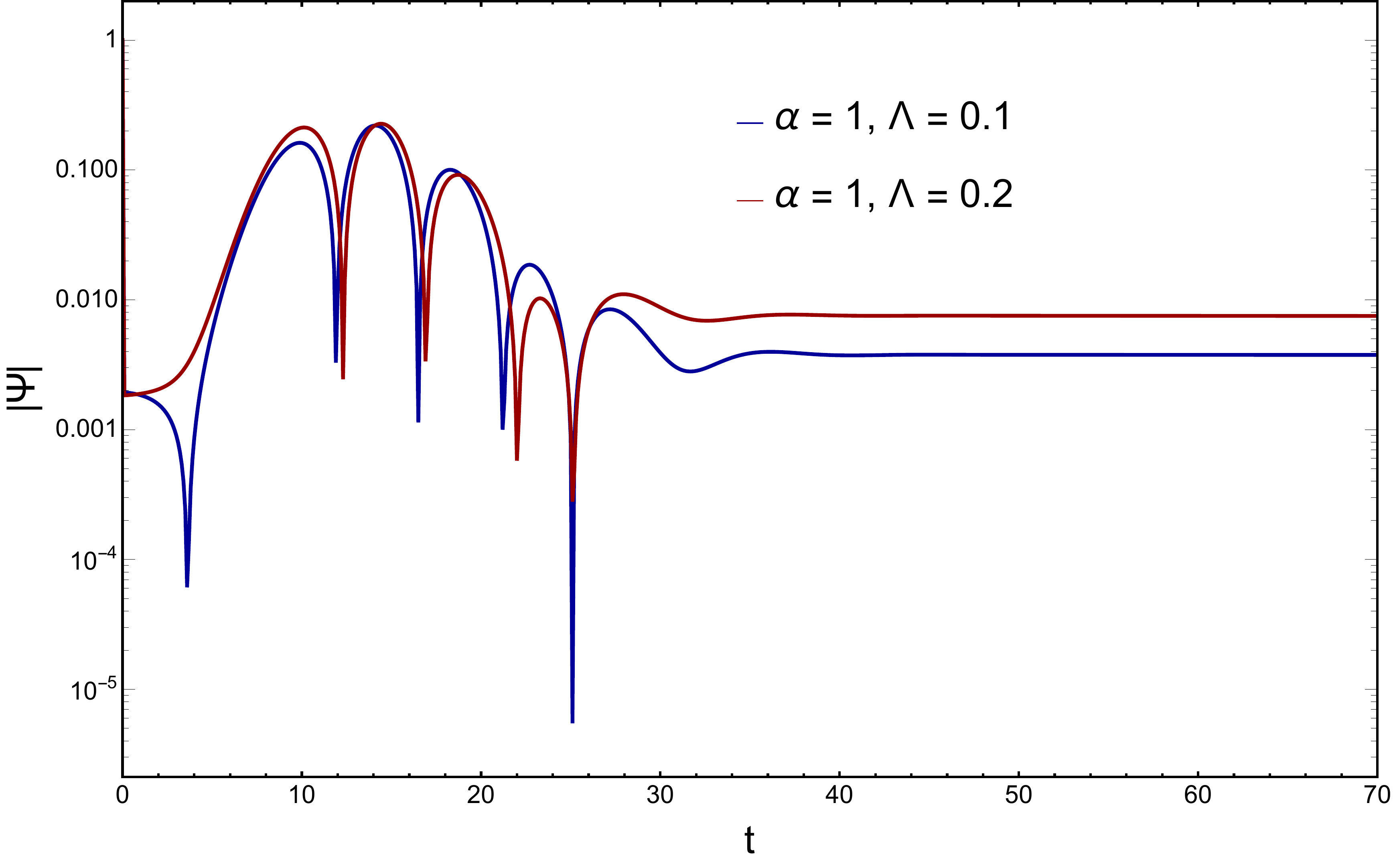} }}    
\qquad
\subfloat{{\includegraphics[scale=0.34]{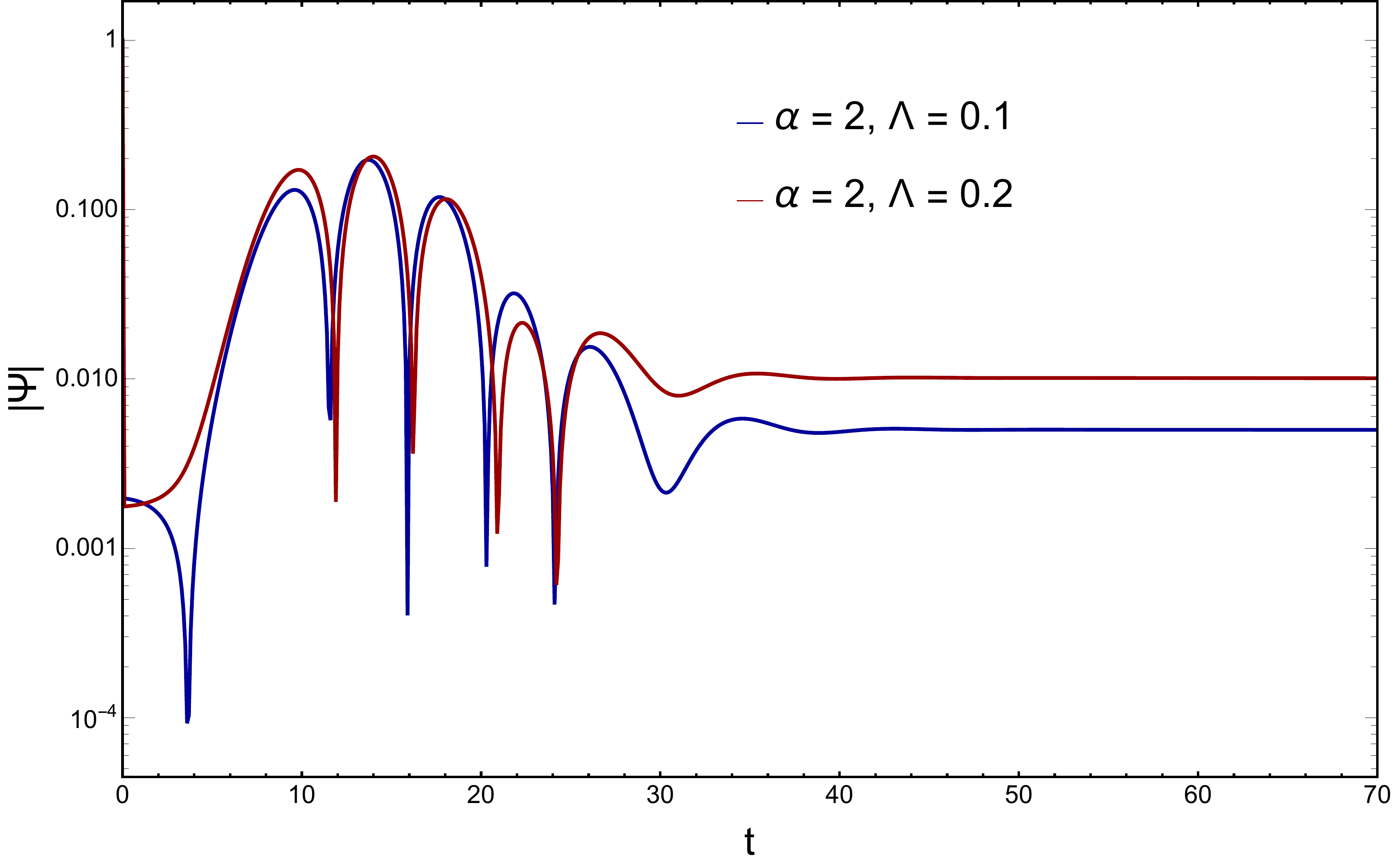} }}
\caption{This figure demonstrates the time domain evolution of of a scalar field on the background of a 4d Gauss Bonnet charged black hole in de Sitter space, confirming an exponentially decaying tail at late times for $\ell=0$ mode. The parameters for these plots chosen are $M=1$ and $(Q/Q_{\rm max})=0.1$. Straight line on a Log plot represents an exponential behaviour. }
 \label{latetime}
\end{figure}

Let us start by studying the dynamics of a massless scalar field on the $4D$ \EGB de Sitter black hole spacetime governed by  \ref{mastereqn}, which can be further expressed as,
 \begin{align}
\left(4 \frac{\partial^2}{\partial u \partial v} + V(u,v) \right)\Phi(u,v)=0
\end{align}
where $u=t-r_*$ and $v=t+r_*$ are the null cone variables. The above equation can be numerically solved in a straight forward manner on a null grid by adapting the following discritization scheme, 
 \begin{align}\label{discrit}
\Phi(u+h,v+h) = -\Phi(u,v)+\Phi(u+h,v)+\Phi(u,v+h)-\frac{h^2}{8}V(u,v)\left[\Phi(u+h,v)+\Phi(u,v+h)\right]
\end{align}
\ref{discrit} allows us to obtain the value of the field $\Phi$ on the entire null grid by starting from an initial data, which we take to be $\Phi(u,v_0)= exp\left(\frac{-(v-12)^2}{10}\right)$. The time domain profile provides the value of field as a function of time i.e., $\Phi(t_0), \Phi(t_0+h), \Phi(t_0+2h)...$etc, which we use to compute the quasinormal mode by a Prony fit algorithm. The time domain profile for a scalar field perturbation on the bachground of $4D$ Gauss Bonnet charged de Sitter black hole is depicted in \ref{latetime} for various choice of parameters. The figures clearly illustrates an exponential late time decay.

\newpage

The scalar quasi-normal modes are eigenfrequencies of the master wave equation \ref{mastereqn} for $s=0$, with the black hole boundary conditions, i.e., purely in-going modes at the event horizon and purely out-going modes at the cosmological horizon,
 \begin{align}
\phi(r\rightarrow r_+)\sim e^{-i\omega r_*}\qquad \text{and} \qquad \phi(r\rightarrow r_c)\sim e^{i\omega r_*}
\end{align}
We also use the well tested Mathematica package developed in Ref.\cite{Jansen:2017oag} and sixth-order WKB approximation for computation of quasinormal frequencies. The numerical value of $\beta = min[-Im(\omega)/\kappa_{-}]$, which determines the violation of \scc\ has been presented in  \ref{table-QNM-GB_4d_scalar} for scalar perturbation.\\
\begin{table}[h]
\begin{centering}
\begin{tabular}{|c|c|c|c|c|c|c|c|}
\hline 
$\alpha$ & $\Lambda$ & $Q/Q_{\textrm{max}}$ & $\ell=0$ & $\ell=1$ & $\ell=10$ & $\ell=10$ (analytical) \tabularnewline
\hline 
\hline 
\multirow{4}{*}{$0.1$} & \multirow{2}{*}{$0.01$} & \multirow{1}{*}{$0.99$}  & $0.83345$ & $0.41607 $ & $0.61264 $& $0.61042 $ \tabularnewline

\cline{3-7} 
 &  & \multirow{1}{*}{$0.995$} & $0.90112$ &  $0.63171$ & $0.91967 $& $0.92069$ \tabularnewline

 \cline{2-7}\cline{3-7}  
 & \multirow{2}{*}{$0.05$} & \multirow{1}{*}{$0.99$} & $0.80343$ &  $1.00052 $ & $0.56873 $& $0.56829 $\tabularnewline
 
\cline{3-7}
 &  & \multirow{1}{*}{$0.995$}  & $0.87892$ & $1.52456 $ & $0.86028$& $0.86224 $ \tabularnewline
 
\cline{7-7}  
\hline 
\multirow{4}{*}{$0.2$} & \multirow{2}{*}{$0.01$} & \multirow{1}{*}{$0.99$}  & $0.86608$ & $0.53995 $ & $0.76877 $& $0.76847$ \tabularnewline

\cline{3-7} 
 &  & \multirow{1}{*}{$0.995$} & $0.92425$ &  $0.80651 $ & $1.14142 $& $1.11378$\tabularnewline

\cline{2-7}\cline{3-7}  
 & \multirow{2}{*}{$0.05$} & \multirow{1}{*}{$0.99$} & $0.88610$ &  $1.28742 $ & $0.75442 $& $0.75292 $ \tabularnewline
 
\cline{3-7}
 &  & \multirow{1}{*}{$0.995$}  & $0.91084$ & $1.93162 $ & $1.05611 $& $1.06065 $ \tabularnewline
 
\cline{7-7} 
\hline 
\multirow{4}{*}{$0.3$} & \multirow{2}{*}{$0.01$} & \multirow{1}{*}{$0.99$}  & $0.90122$ & $0.68265 $ & $0.94530 $& $0.93816 $ \tabularnewline

\cline{3-7} 
 &  & \multirow{1}{*}{$0.995$} & $0.94164$ &  $1.00844 $ & $1.39002 $& $1.38595 $ \tabularnewline

\cline{2-7}\cline{3-7}  
 & \multirow{2}{*}{$0.05$} & \multirow{1}{*}{$0.99$} & $0.91747$ &  $1.61862 $ & $0.86334 $& $0.86240 $ \tabularnewline
 
\cline{3-7}
 &  & \multirow{1}{*}{$0.995$}  & $0.93268$ & $2.40133 $ & $1.28835 $& $1.27295 $ \tabularnewline
\cline{7-7} 
\hline 
\end{tabular}
\par\end{centering}
\caption{We present the numerical values of $\beta\equiv \{-\textrm{min.}~(\textrm{Im}~\omega_{n})/\kappa_{\rm ch}\}$ in the first three column($\ell=0$, $\ell=1$, $\ell=10$) for various choice of Gauss-Bonnet coupling constant ($\alpha$), cosmological constant $\Lambda$ and electric charge $(Q/Q_{\rm max})$ for $M=1$. 
The last column represents the value of $\beta$ for the photon sphere modes computed from the corresponding analytical expression presented in \ref{Section 3}. The numerical and analytical value of $\beta$ is clearly in close agreement.
}
\label{table-QNM-GB_4d_scalar}
\end{table}

\subsection{Validity of strong cosmic censorship}
\subsubsection{Scalar perturbation}
\begin{figure}[!htp]
\centering
\subfloat{{\includegraphics[scale=0.37]{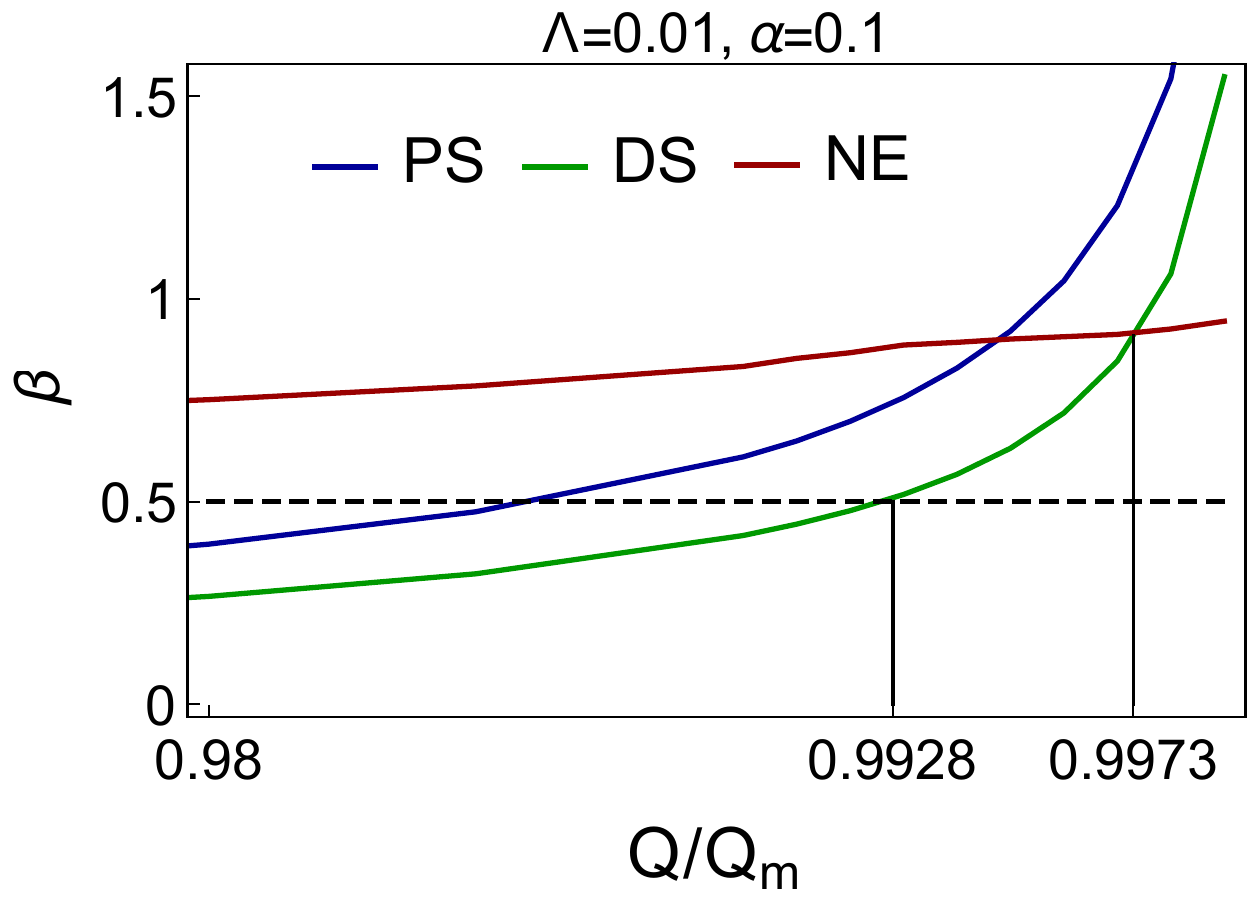} }}    
\qquad
\subfloat{{\includegraphics[scale=0.37]{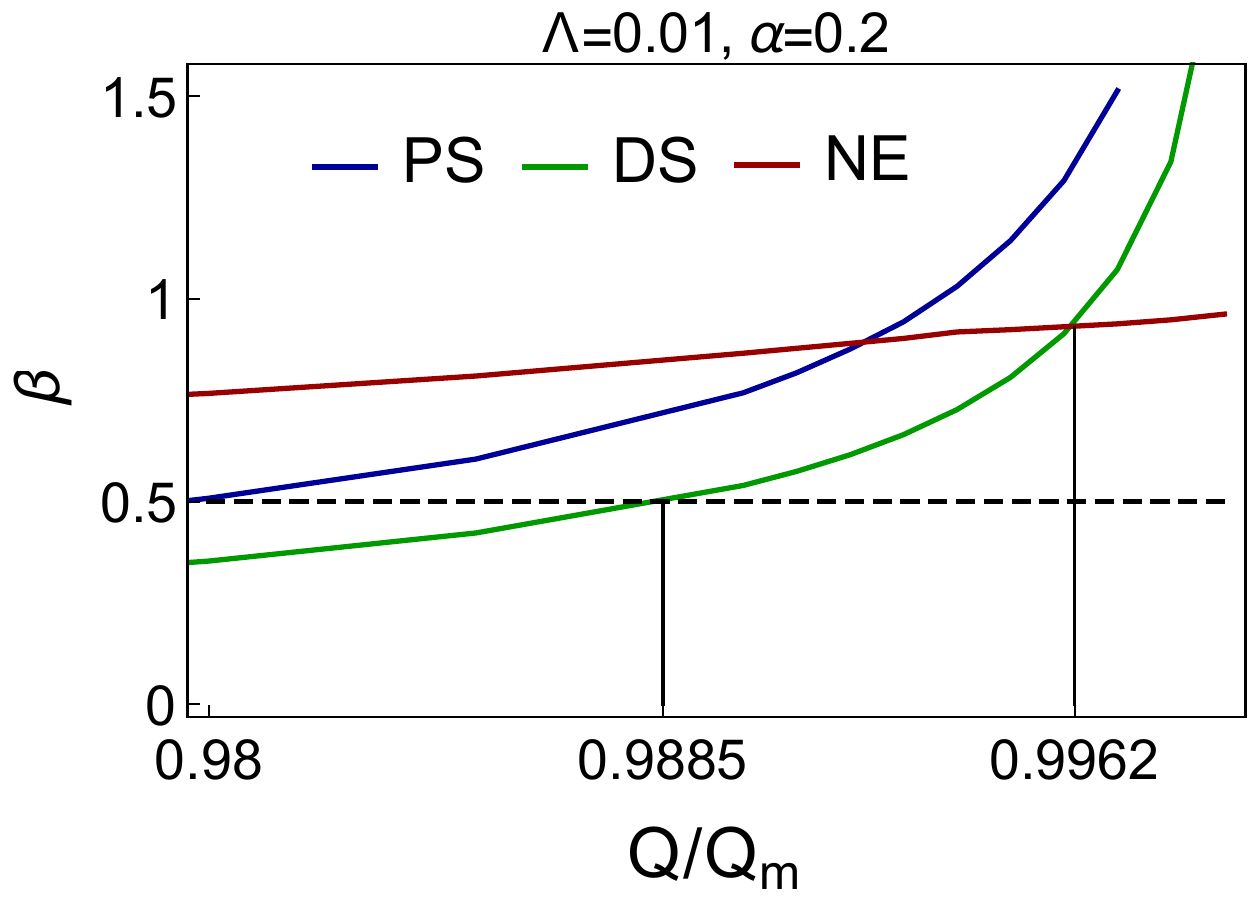} }}
\qquad
\subfloat{{\includegraphics[scale=0.39]{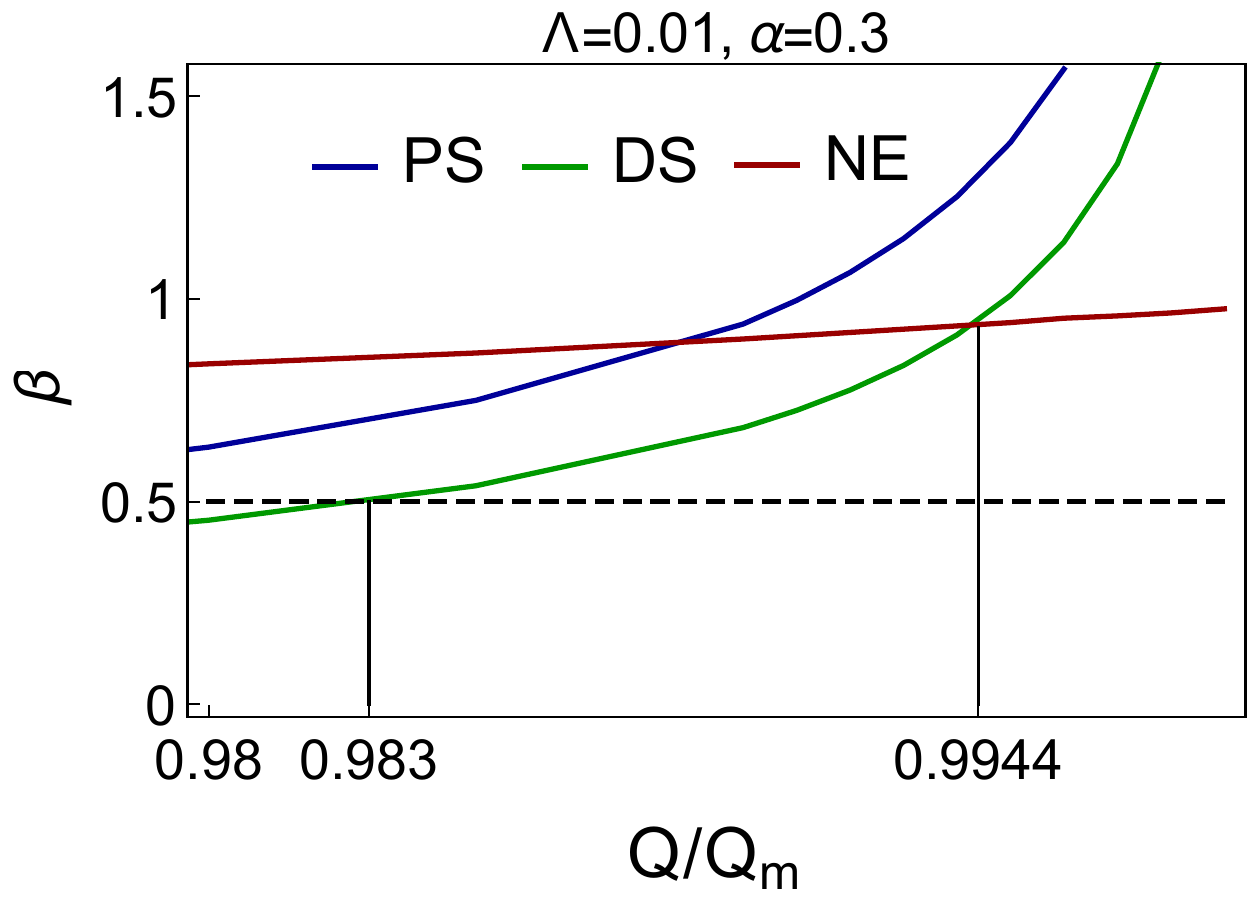} }}
\qquad
\subfloat{{\includegraphics[scale=0.37]{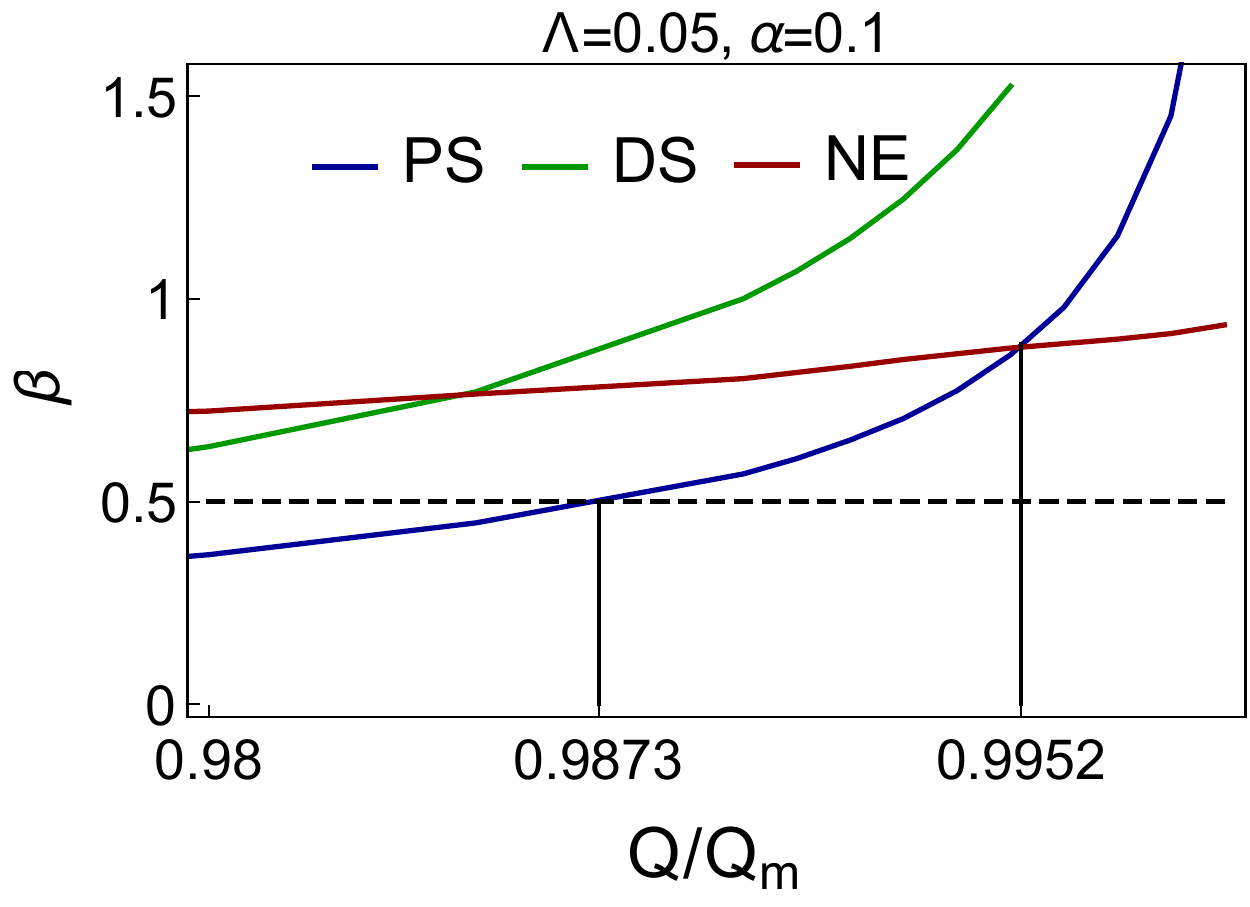} }}
\qquad
\subfloat{{\includegraphics[scale=0.37]{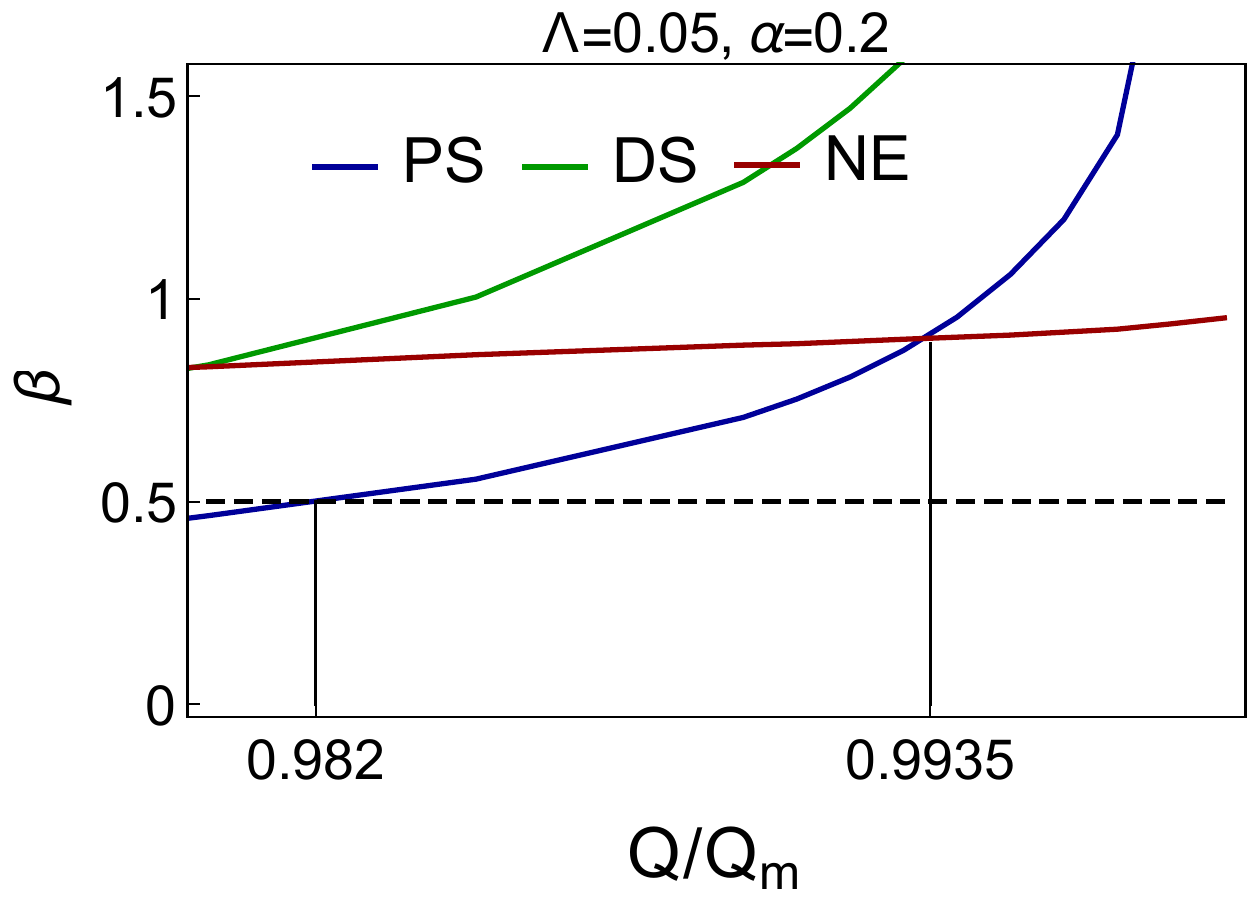} }}
\qquad
\subfloat{{\includegraphics[scale=0.39]{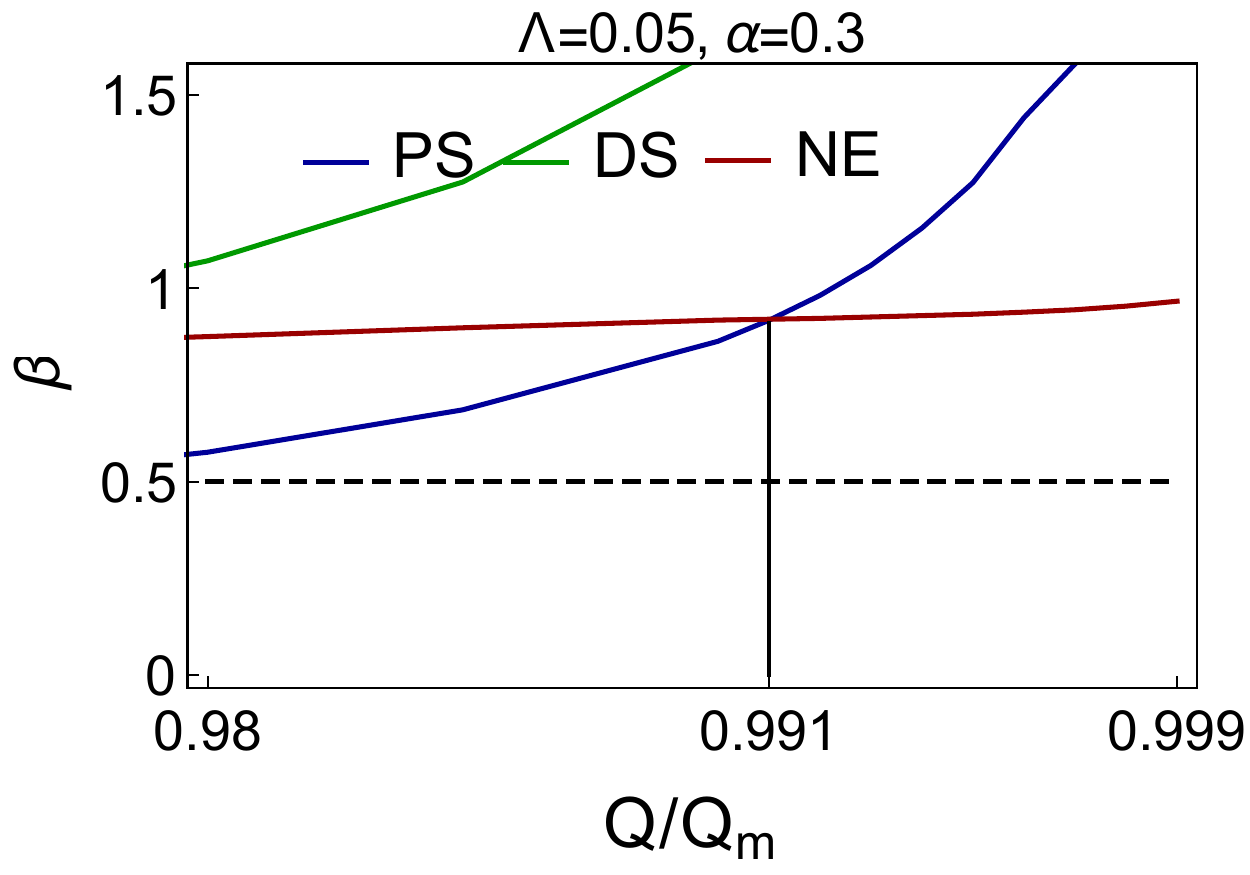} }}
\qquad
\subfloat{{\includegraphics[scale=0.37]{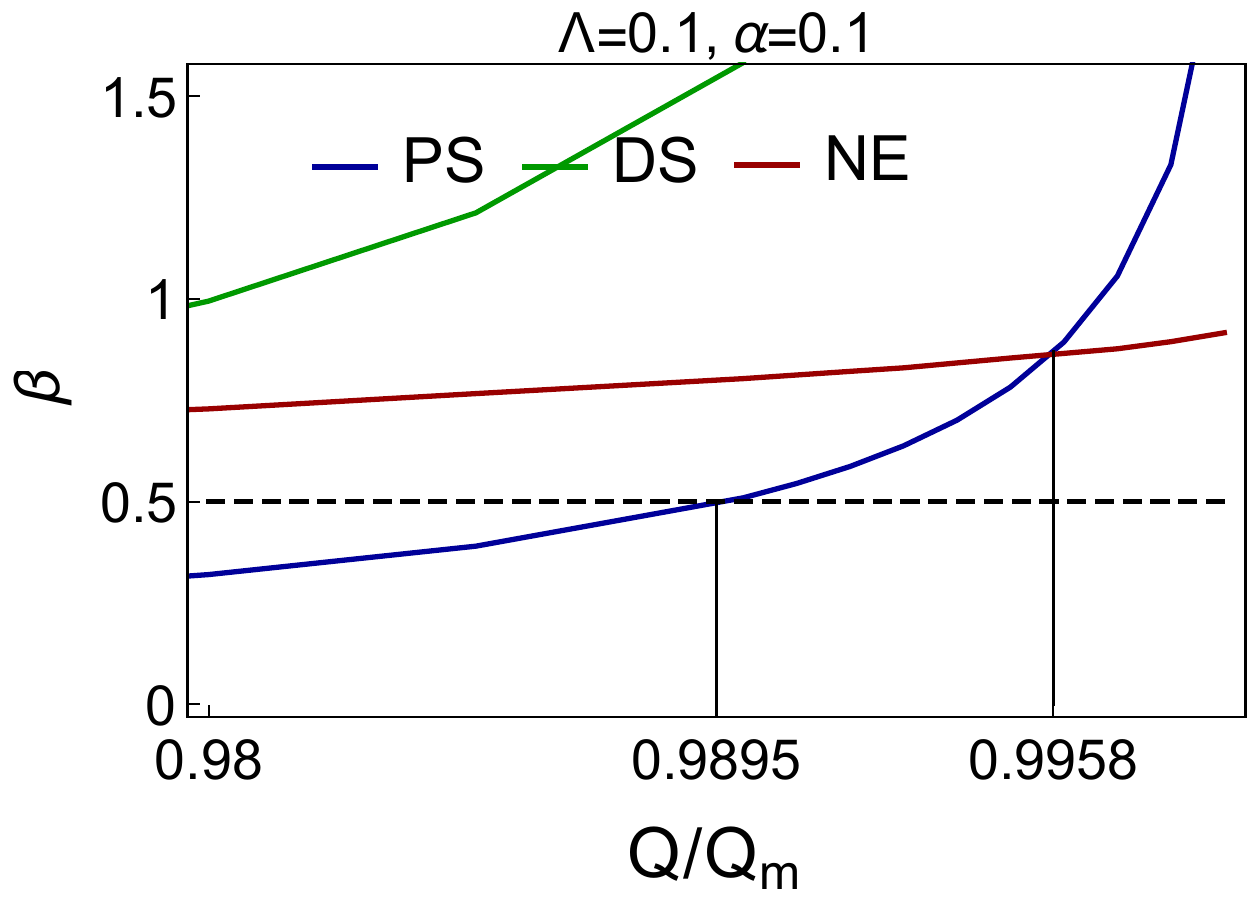} }}
\qquad
\subfloat{{\includegraphics[scale=0.37]{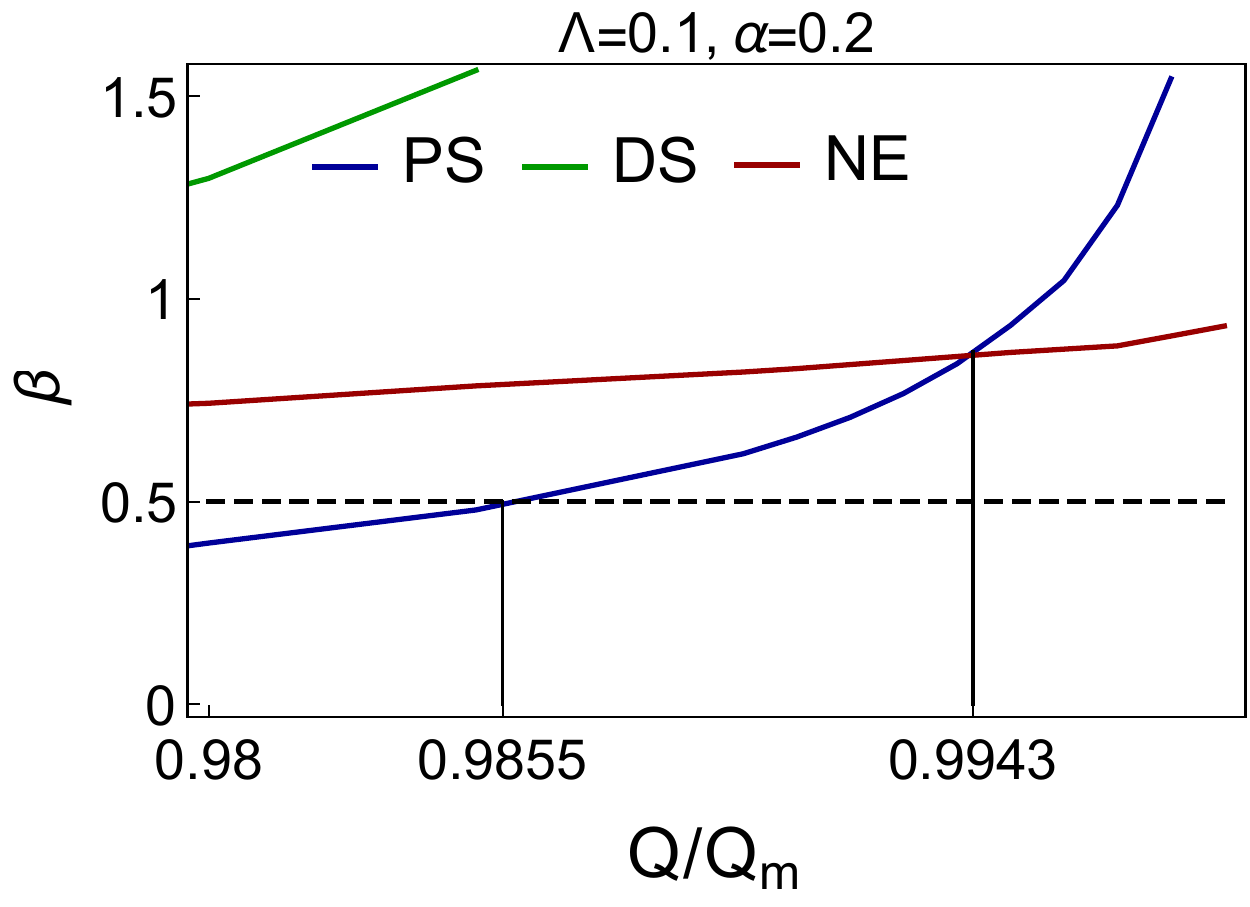} }}
\qquad
\subfloat{{\includegraphics[scale=0.39]{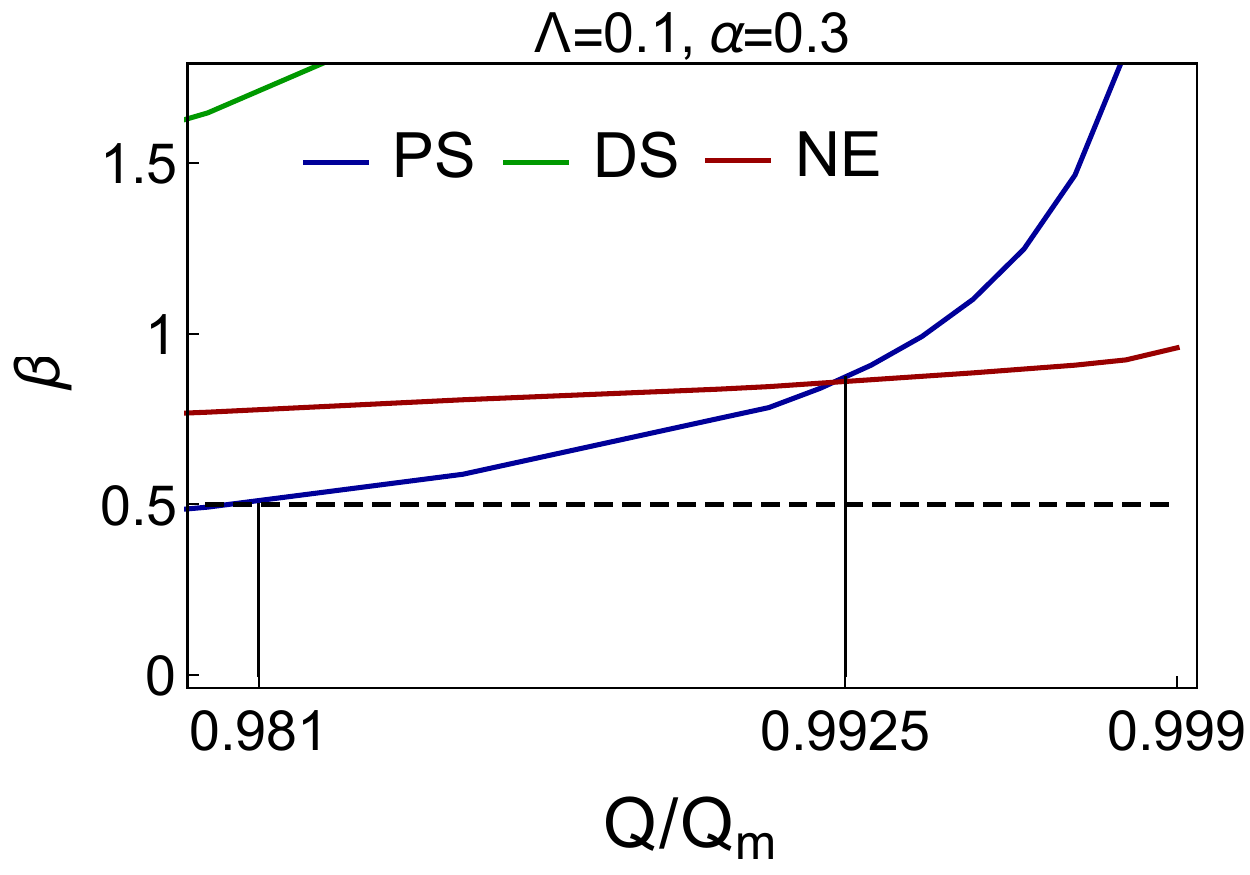} }}
\caption{This figure demonstrates the variation of $\beta$ with the ratio $(Q/Q_{\rm m})$ for all three family of modes, namely the \PS modes(blue), the \DS modes(green) and the \NE modes(red) for various choice of cosmological constant($\Lambda$) and the Gauss-Bonnet coupling constant($\alpha$). The first vertical line in these plots corresponds to the value of $Q/Q_{\rm m}$ for which the \scc\ is violated for the first time. In contrast, the second vertical line represents the value of $Q/Q_{\rm m}$, where the near extremal mode becomes dominant. Mass of the black hole is taken to be unity.}
 \label{4d_GB_scalar}
\end{figure}

Having obtained the necessary numerical results, we are now all set to study the validity of \scc\ for charged de Sitter black holes in \EGB\  gravity for scalar field perturbation. To that end, we plot the variation of $\beta$ for all three families of modes, with respect to $(Q/Q_{\rm m})$ in \ref{4d_GB_scalar}, where $Q_{m}$ represents the extremal value of the black hole charge. The \PS\, \DS\, and \NE\ modes are depicted by blue, green and red lines respectively. 
The de Sitter mode appears to dominate the \qnm\ spectrum for a small value of the cosmological constant, while the \NE\ modes are dominating for a larger value of the extremal charge $Q_{\rm m}$. As evident from \ref{4d_GB_scalar}, violation of the \scc\ conjecture occurs with respect to scalar perturbation in the near extremal region of a charged de Sitter black hole in the $4D$ \EGB\ theory of gravity. Moreover, $\beta$ for the dominating mode never exceed unity, suggesting that the scalar fields can be extended beyond the \ch\ with regularity not smoother than $H^{1}_{loc}$. Furthermore, the effect of Gauss-Bonnet coupling constant on the violation of \scc\ can be realized from \ref{table-QNM-GB_4d_scalar}. The $\beta$ value for the dominant mode appears to be increasing with the increasing strength of coupling $\alpha$, suggesting that the violation of \scc\ is stronger for the larger value of the coupling. 

\newpage
\subsubsection{Electromagnetic perturbation}
Having studied the case of scalar perturbation in the previous sub-section, now we turn our attention to electromagnetic perturbation. 
The \qnm\ mode spectrum is governed by the master equation \ref{mastereqn} with spin $s=1$. By following an identical approach, we first compute the \qnm\ frequencies by imposing the black hole boundary conditions and then proceed further to plot $\beta$ with respect to $(Q/Q_m)$ in order to check the validity of \scc. Note that, the \qnm\ frequency spectrum for electromagnetic perturbation starts with the $\ell=1$ mode. The numerical value for $\beta$ is presented in \ref{table-QNM-GB_4d_em} .

\begin{table}[h]
\begin{centering}
\begin{tabular}{|c|c|c|c|c|c|}
\hline 
$\alpha$ & $\Lambda$ & $Q/Q_{\textrm{max}}$ & $\ell=1$ & $\ell=10$  \tabularnewline
\hline 
\hline 
\multirow{4}{*}{$0.1$} & \multirow{2}{*}{$0.01$} & \multirow{1}{*}{$0.99$}  & $0.83515$ & $0.60985 $  \tabularnewline

\cline{3-5} 
 &  & \multirow{1}{*}{$0.995$} & $1.26678$ &  $0.91845$  \tabularnewline

 \cline{2-5}\cline{3-5}  
 & \multirow{2}{*}{$0.05$} & \multirow{1}{*}{$0.99$} & $2.02339$ &  $0.56784 $\tabularnewline
 
\cline{3-5}
 &  & \multirow{1}{*}{$0.995$}  & $3.08332$ & $0.85891 $  \tabularnewline
 
\cline{5-5}  
\hline 
\multirow{4}{*}{$0.2$} & \multirow{2}{*}{$0.01$} & \multirow{1}{*}{$0.99$}  & $1.08277$ & $0.76770 $  \tabularnewline

\cline{3-5} 
 &  & \multirow{1}{*}{$0.995$} & $1.61730$ &  $1.14001 $ \tabularnewline

\cline{2-5}\cline{3-5}  
 & \multirow{2}{*}{$0.05$} & \multirow{1}{*}{$0.99$} & $2.60385$ &  $0.70718 $ \tabularnewline
 
\cline{3-5}
 &  & \multirow{1}{*}{$0.995$}  & $3.90682$ & $1.05437 $  \tabularnewline
 
\cline{5-5} 
\hline 
\multirow{4}{*}{$0.3$} & \multirow{2}{*}{$0.01$} & \multirow{1}{*}{$0.99$}  & $1.36892$ & $0.94390 $  \tabularnewline

\cline{3-5} 
 &  & \multirow{1}{*}{$0.995$} & $2.02232$ &  $1.40167 $  \tabularnewline

\cline{2-5}\cline{3-5}  
 & \multirow{2}{*}{$0.05$} & \multirow{1}{*}{$0.99$} & $3.27373$ &  $0.86188 $ \tabularnewline
 
\cline{3-5}
 &  & \multirow{1}{*}{$0.995$}  & $4.85672$ & $1.27599 $ \tabularnewline
\cline{5-5} 
\hline 
\end{tabular}
\par\end{centering}
\caption{Numerical value of the $\beta\equiv \{-\textrm{min.}~(\textrm{Im}~\omega_{n})/\kappa_{\rm ch}\}$ has been presented for the \DS\ modes($\ell=1)$ and \PS\ modes($\ell=10$) for various choice of Gauss-Bonnet coupling constant ($\alpha$), cosmological constant $\Lambda$ and electric charge $(Q/Q_{\rm max}$. Mass parameter(M) of the black hole is taken to be unity.
}
\label{table-QNM-GB_4d_em}
\end{table}

\begin{figure}[!htp]
\centering
\subfloat{{\includegraphics[scale=0.37]{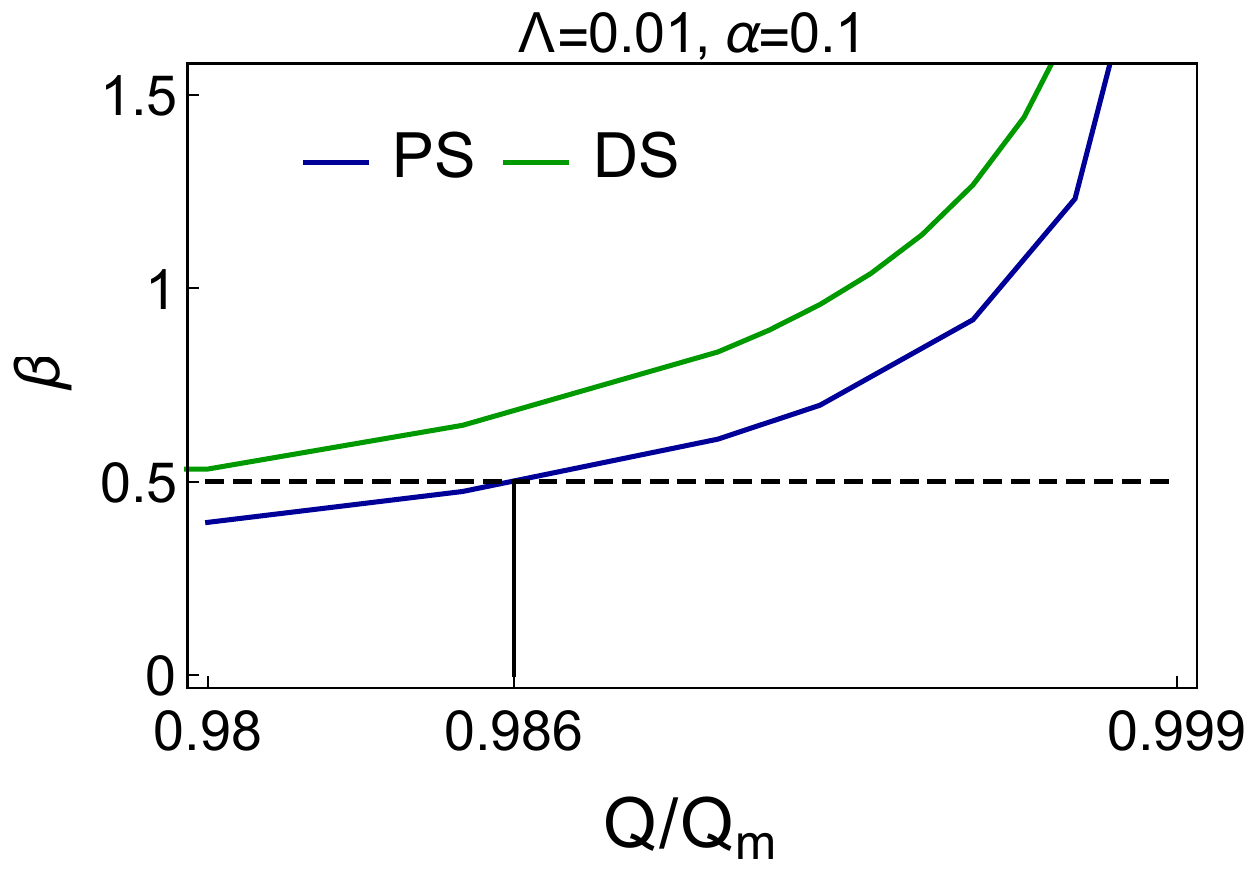} }}    
\qquad
\subfloat{{\includegraphics[scale=0.37]{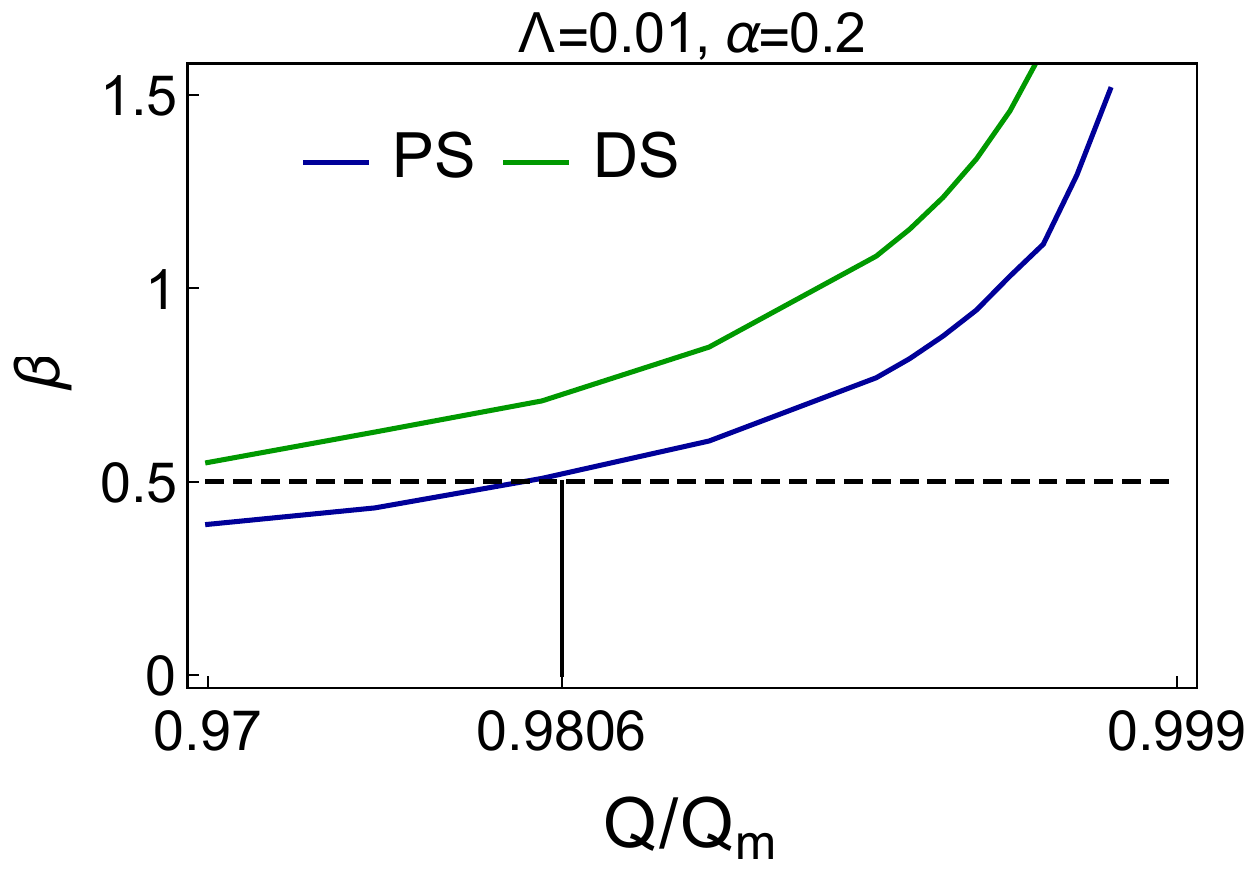} }}
\qquad
\subfloat{{\includegraphics[scale=0.39]{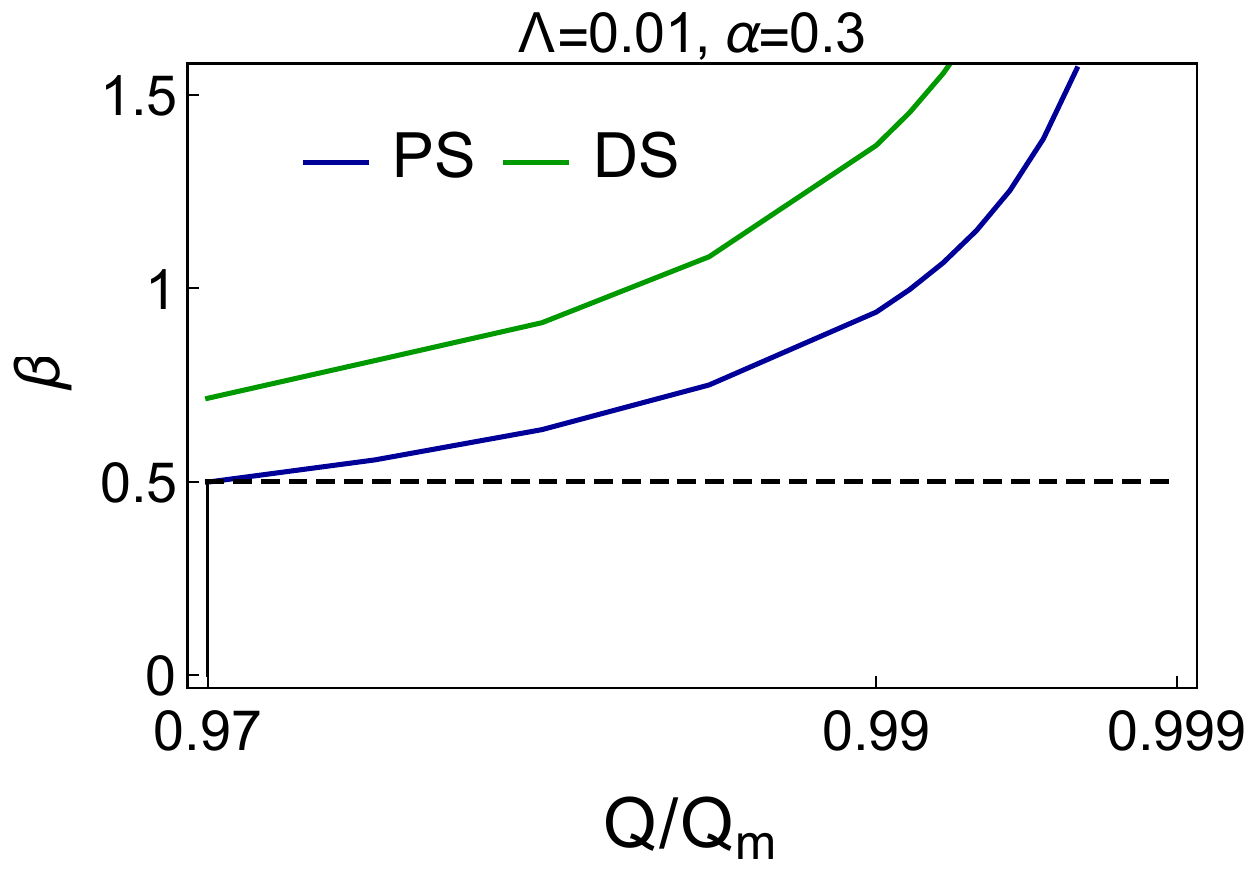} }}
\qquad
\subfloat{{\includegraphics[scale=0.37]{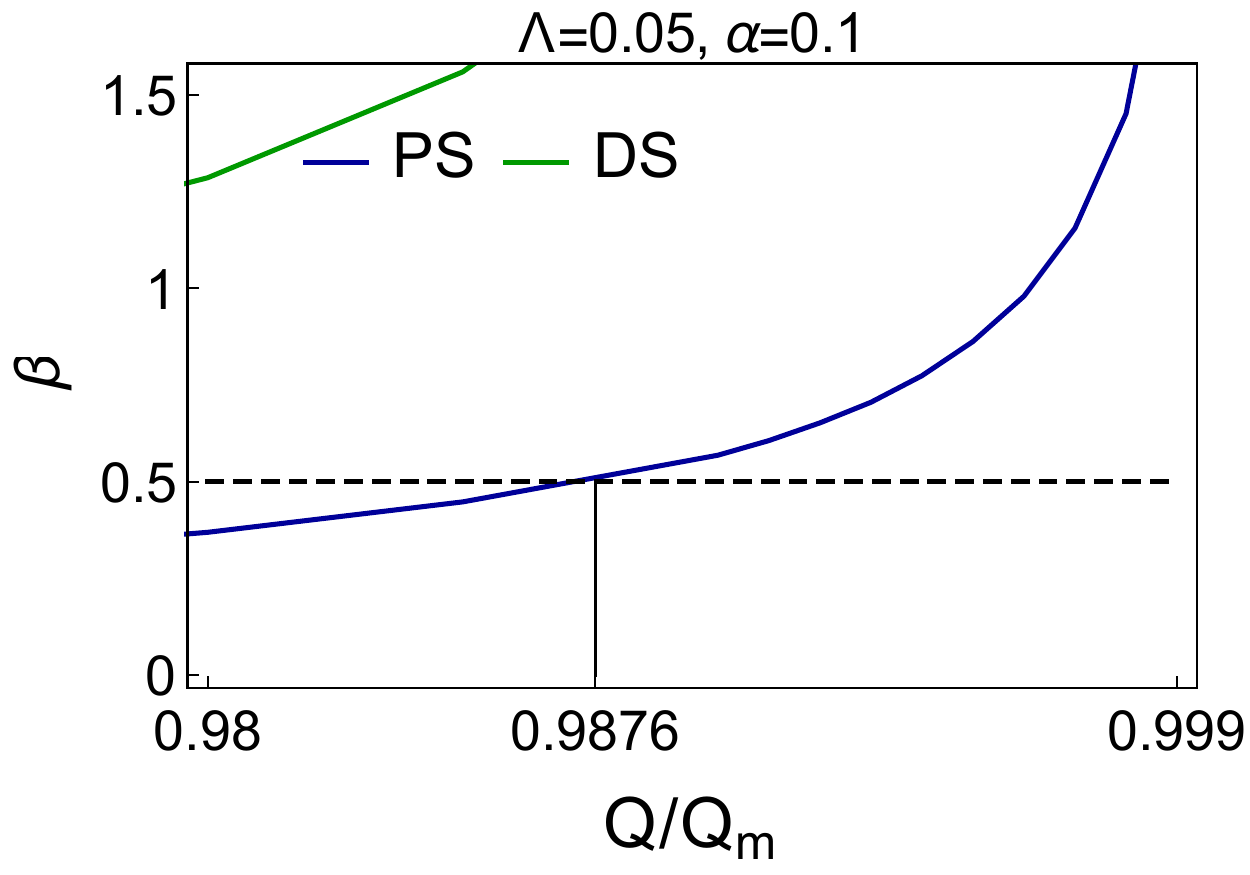} }}
\qquad
\subfloat{{\includegraphics[scale=0.37]{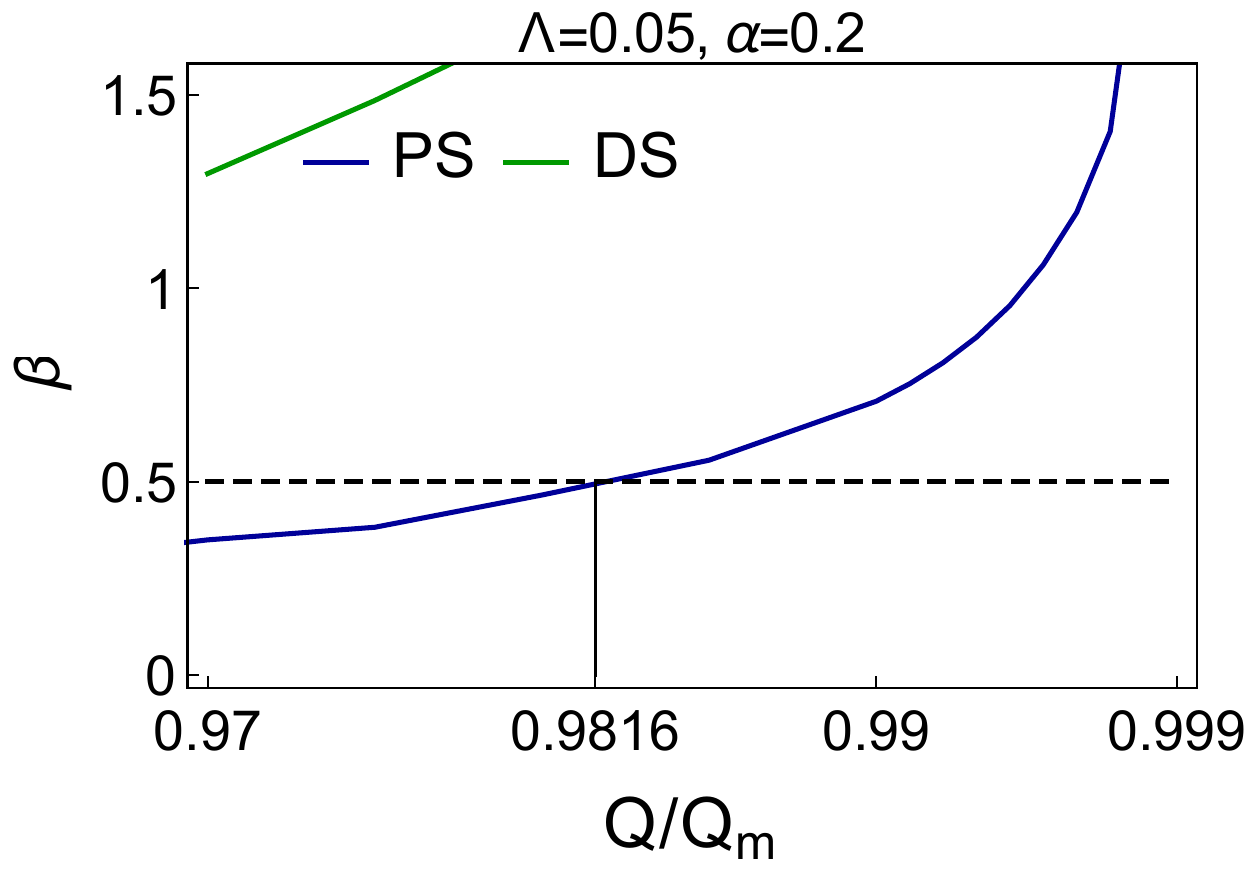} }}
\qquad
\subfloat{{\includegraphics[scale=0.39]{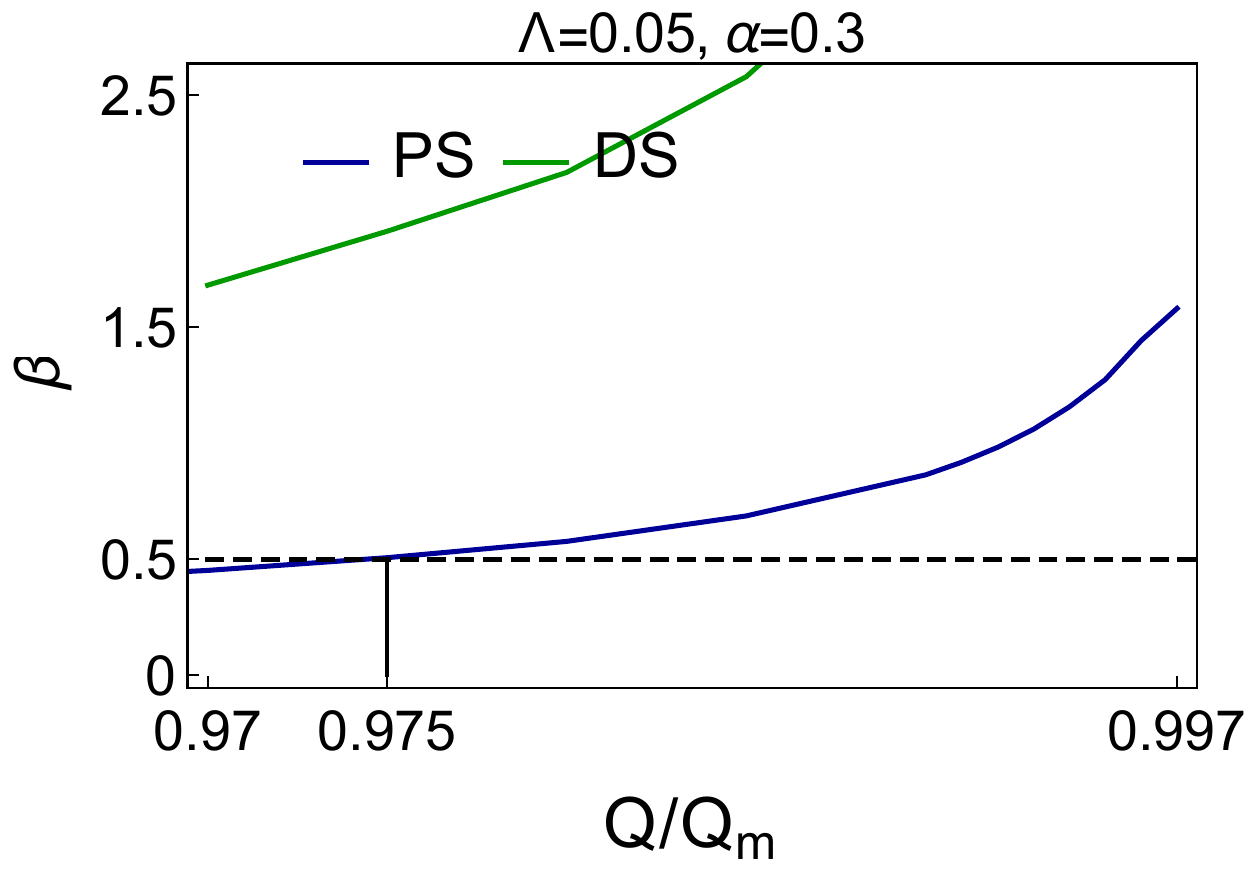} }}
\qquad
\subfloat{{\includegraphics[scale=0.37]{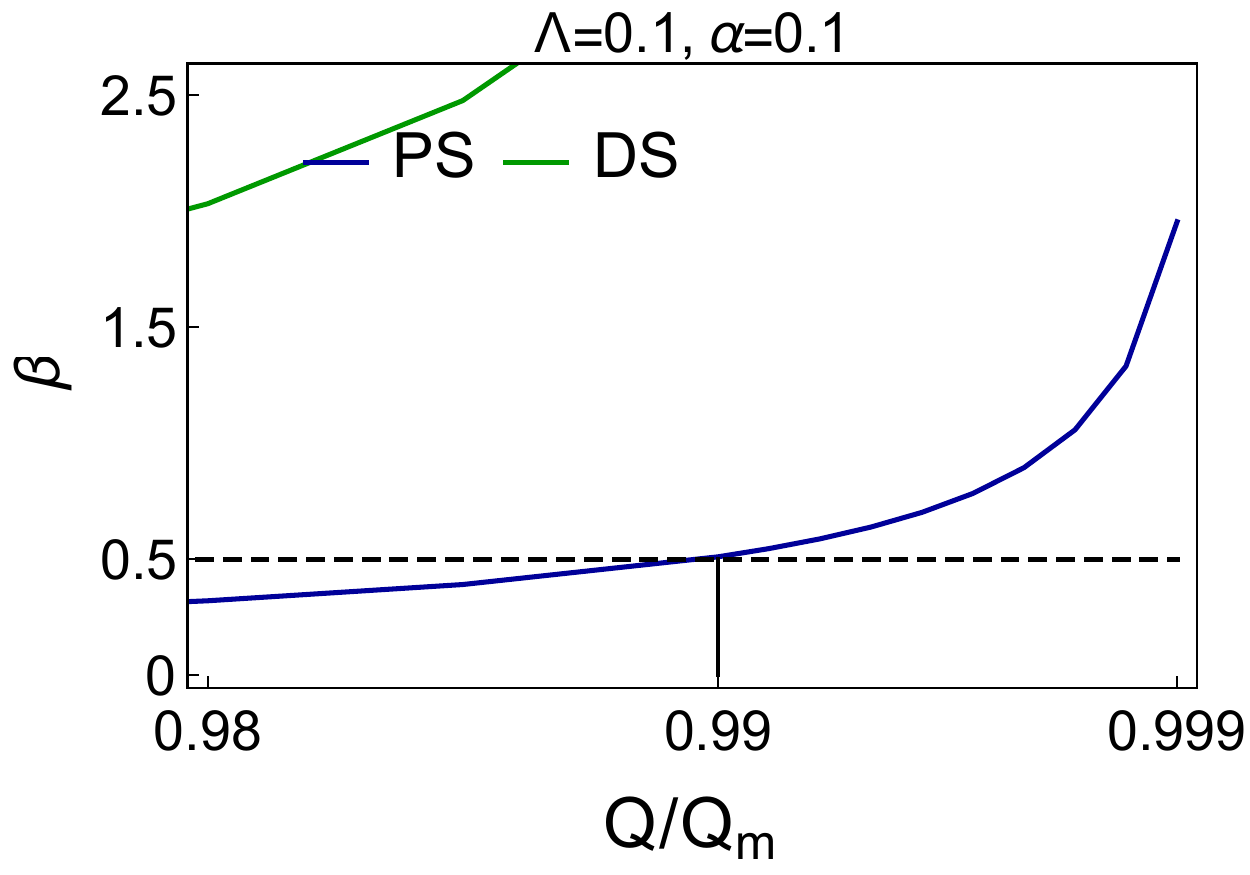} }}
\qquad
\subfloat{{\includegraphics[scale=0.37]{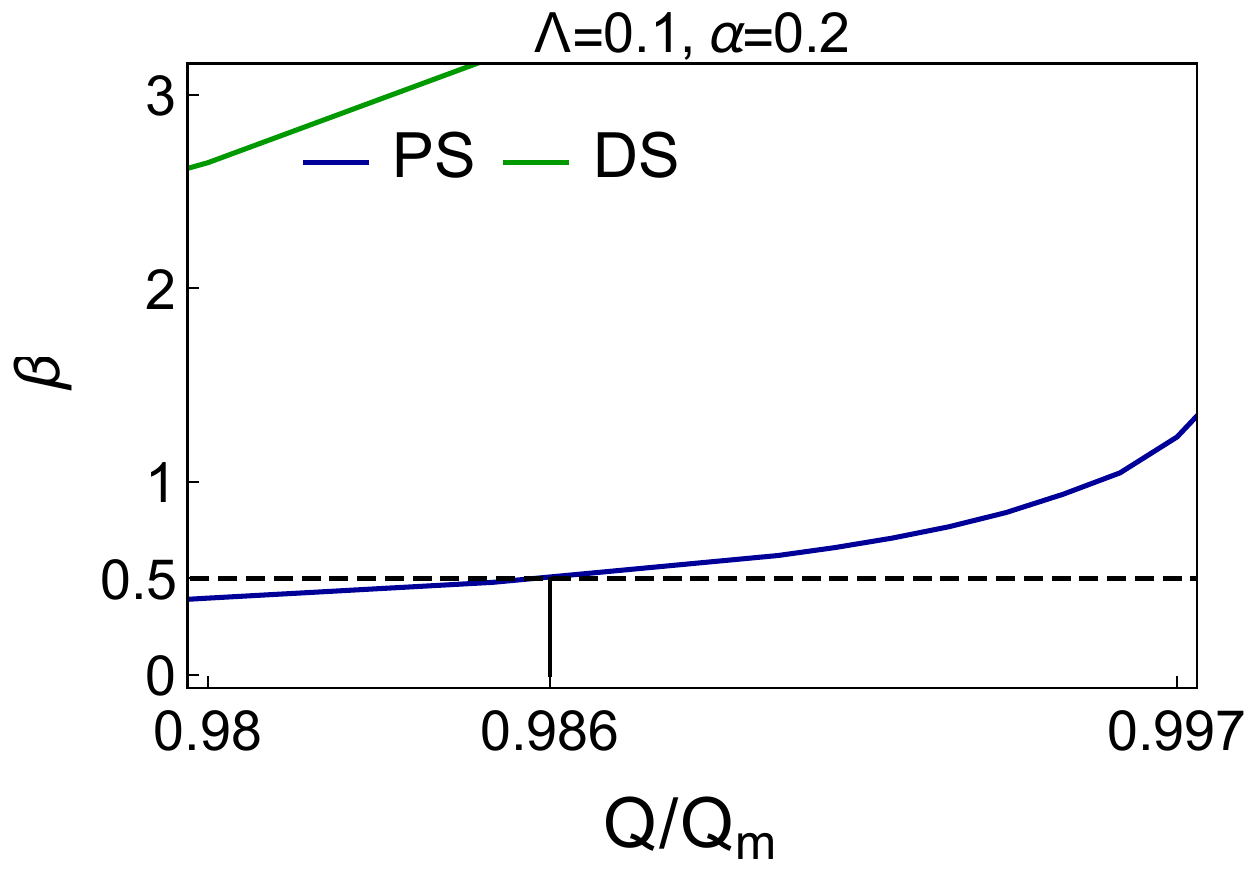} }}
\qquad
\subfloat{{\includegraphics[scale=0.39]{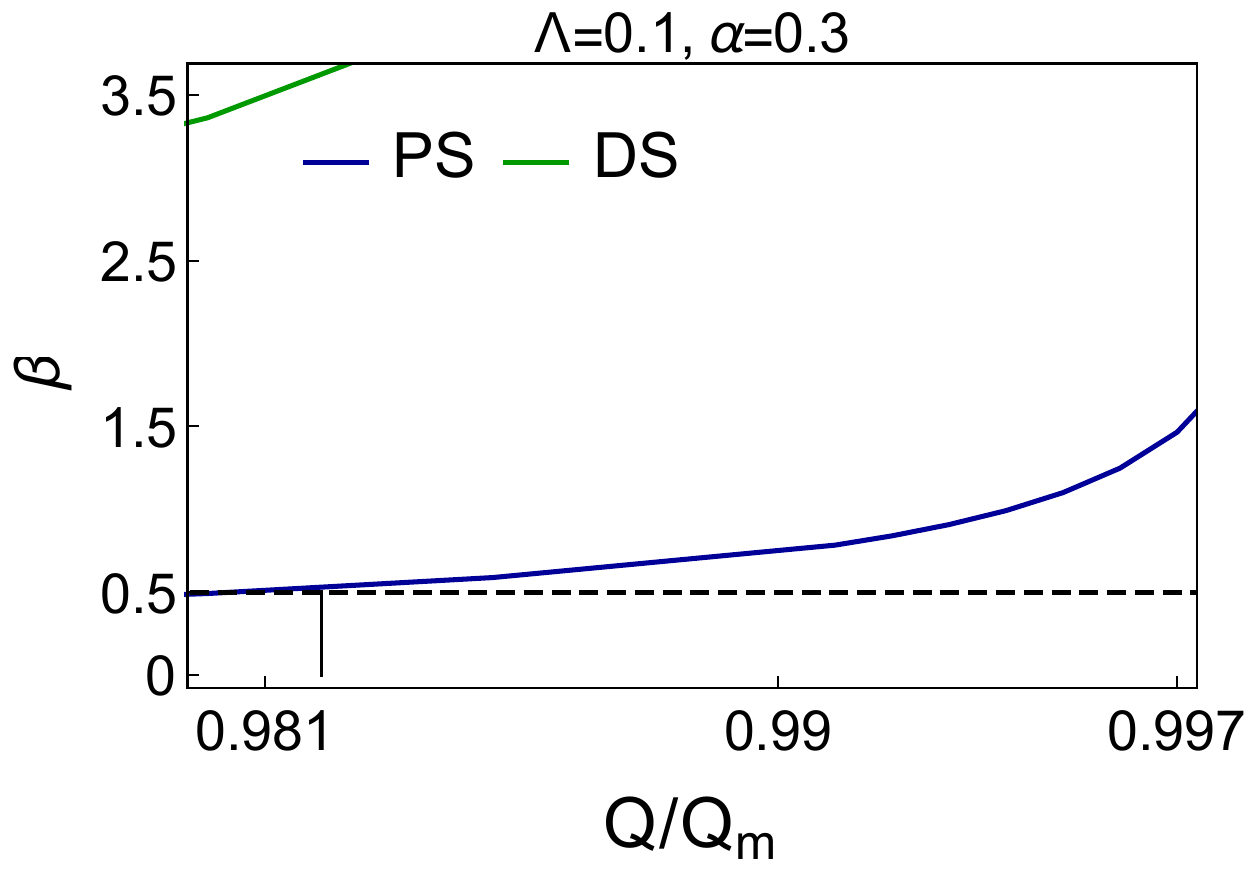} }}
\caption{The parameter $\beta$ is presented corresponding to electromagnetic perturbation for the \DS modes(green) and the \PS modes(blue) for various choice of cosmological constant($\Lambda$) and the Gauss-Bonnet coupling constant($\alpha$). The vertical lines in these plots correspond to the value of $Q/Q_{\rm m}$ for which the \scc\ is violated for the first time, The Mass parameter of the black hole is taken to be unity.
}
 \label{4d_GB_em}
\end{figure}

In \ref{4d_GB_em} we demonstrate the interplay between the de Sitter(green) and photon sphere(blue) modes for electromagnetic perturbations for various choice of parameters. We observe that the Christodulo version of the \scc\ conjecture is violated in the near extremal region even with respect to electromagnetic perturbations. Moreover, for fixed cosmological constant and charge($Q/Q_{\rm m}$),  the value of $\beta$ corresponding to the dominant mode appears to increase for increasing strength of the coupling constant. Unlike the case of the scalar field, the $\beta$ value for the electromagnetic perturbation corresponding to the dominant mode exceeds unity in the near extremal region. This further suggests that the electromagnetic perturbations turns out to be $C^{1}$ extendible across the \ch, leading to an even stronger violation of the conjecture.

\newpage
\section{Discussions}\label{Section 5}
In recent years several counter-example to the \scc\ is found, and it has been realized that a violation of the conjecture occurs in the near extremal region of charged de Sitter black holes, suggesting \gr\ to be not deterministic. 
However, general relativity, being a non-renormalizable theory, must be supplemented by higher curvature correction terms in strong gravity regime. Therefore it is essential to study the fate of \scc\ in the presence of such higher-dimensional operators.
Among a wide class of such theories of gravity, the \EGB\ theory and its Lovelock generalization play a very crucial role. It is well known that the Gauss-Bonnet term in four dimensions turns out to be topological and only contribute to the dynamics of gravitational action when $D>4$. However, recently the \EGB\ theory has been reformulated in $4D$ as the $D\to 4$ limit of the higher dimensional theory after rescaling the coupling constant as $\alpha\to\alpha/(D-4)$. The regularized $4D$ \EGB\ theory represents the only higher curvature theory in four dimensions with field equation containing at most up to the second derivative of the metric. This has been of great interest and lead to several interesting works recently. \\

In this article, we have studied the validity of \scc\ for charged de Sitter black hole in the context of \EGB\ theory in $4D$. 
By taking into account, linear scalar and electromagnetic perturbation, we have explicitly demonstrated that a violation of the conjecture occurs in the near extremal region. 
We have started by computing the quasi-normal frequencies numerically and subsequently studied the variation of $\beta$ with respect to $(Q/Q_{\rm m})$. Furthermore, we have studied the late time dynamics of scalar field on the charged de Sitter solution of this theory and established the expected exponential tail. The numerical data for both scalar and electromagnetic perturbation are presented in \ref{table-QNM-GB_4d_scalar} and  \ref{table-QNM-GB_4d_em} respectively. For scalar perturbation(\ref{4d_GB_scalar}) we tested the \scc\ by considering the effect of all three family of dominant modes namely, the \PS($\ell=10$), \DS($\ell=1$) and \NE($\ell=0$) modes. However, since for the electromagnetic perturbation the $\ell=0$ mode doesn't contribute to the quasi-normal spectrum, we studied the variation of $\beta$ with respect to \DS($\ell=1$) and \PS($\ell=10$) modes. From this analysis, we conclude that a violation of the \scc\ occurs near the extremal region of a charged de Sitter black hole in the novel $4D$ \EGB\ theory of gravity. The \scc\ conjecture in the context of electromagnetic perturbation has not been studied before. We have also demonstrated that, for electromagnetic perturbation the value of $\beta$ exceeds unity for certain range of parameters leading to a violation of the $C^{1}$ version of \scc\ conjecture along with the Christodoulou's version of the conjecture. However for the scalar perturbations value of $\beta$ never exceeds unity. This indicate that, the violation of \scc\ turns out to be stronger for electromagnetic perturbations.\\

We have also studied the effect of Gauss-Bonnet coupling constant on the violation of \scc. Identical to the higher-dimensional \EGB\  and pure Lovelock case, the violation of \scc\ appears to be stronger with increasing strength of coupling in this theory. This further has lead us to the conclusion that the effect of coupling on the violation of \scc\ conjecture must be a general feature of any higher curvature theory rather than the perturbation itself. Comparing with our earlier work\cite{Mishra:2019ged}, we find that the violation of \scc\ seems to be stronger in higher dimensional spacetimes. As an extension of our analysis one can consider the case of gravitational perturbations. Since the bound on $\beta$ is different, it would be an interesting exercise to study whether the \scc\ can be rescued for gravitational perturbation.
\section*{Acknowledgement}

The author would like to thank Sudipta Sarkar, Sumanta Chakraborty and Rajes Ghosh for many helpful discussions. 

\bibliography{SCC}

\bibliographystyle{./utphys1}

\end{document}